\documentclass[lettersize,journal]{IEEEtran}
\usepackage{amsmath,amsfonts}
\usepackage{algpseudocode}
\usepackage{algorithm}[H]
\usepackage{algorithmicx}
\usepackage{array}
\usepackage[caption=false,font=normalsize,labelfont=sf,textfont=sf]{subfig}
\usepackage{textcomp}
\usepackage{stfloats}
\usepackage{url}
\usepackage{verbatim}
\usepackage{graphicx}
\usepackage{cite}
\usepackage[]{caption, subfig} %
\usepackage{booktabs} %
\usepackage{diagbox}  %
\usepackage{amssymb}  %
\usepackage{amsthm}
\usepackage{siunitx}
\usepackage{mathrsfs}
\usepackage[acronym, toc]{glossaries}
\usepackage{soul}
\usepackage[usenames,dvipsnames]{xcolor}
\usepackage{outlines}
\usepackage{pgfplots}
\usepgfplotslibrary{
    external,
    statistics,
}
\usetikzlibrary{
	external,
	calc,
	patterns,
	decorations.pathmorphing,
	decorations.markings,
	decorations.text,
	positioning,
	backgrounds,
	automata,
	arrows.meta,
	shapes,
	arrows,
	pgfplots.fillbetween,
	shadows,
	fadings
}
\pgfplotsset{compat=1.18}

\algrenewcommand\algorithmicrequire{\textbf{Input:}}
\algrenewcommand\algorithmicensure{\textbf{Output:}}

\tikzset{
	pics/vessel/.style args={#1}{ %
		code={

			\draw [color=#1, line width=1pt, fill=#1] (-0.5,-1) -- (-0.5,1) -- (0, 1.5) --(0.5,1) -- (0.5,-1) -- cycle;

		}
	}
}

\tikzset{
	pics/ownShipRegions/.style args={#1,#2,#3}{ %
		code={
			\node[](#1) at (0,0){};
			
			\draw [fill=orange!30] (0 : #3) arc[start angle=0,delta angle=360,radius=#3] -- cycle;
			\draw[fill=red!30 ] (#1) -- (100 : #3) arc[start angle=100,delta angle=112.5-10,radius=#3] -- cycle;
			\draw[fill=blue!30 ] (#1) -- (100 : #3) arc[start angle=100,delta angle=-20,radius=#3] -- cycle;
			\draw[fill=green!30 ] (#1) -- (80 : #3) arc[start angle=80,delta angle=-(112.5-10),radius=#3] -- cycle;
			
			\draw[-,very thick](0,0.5)--(0,0.5+#2);
		}
	}
}

\begin{document}

\newcommand{\pnha}[2]{{\color{Red}{NH: #1: \textit{#2}}}}
\newcommand{\dimpa}[1]{{\color{RoyalBlue}{DP: #1}}}
\newcommand{\roga}[1]{{\color{Green}{RG: #1}}}
\newcommand{\mobl}[1]{{\color{Orange}{MB: #1}}}

\newcommand{\rebuttal}[1]{{\color{red}{#1}}}

\newcommand{\figref}[1]{Fig.~\ref{#1}}

\newtheorem{defn}{Definition}
\newtheorem{definition}{Definition}
\newtheorem{rem}[defn]{Remark}
\newtheorem*{remark}{Remark}
\newtheorem{lem}[defn]{Lemma}
\newtheorem{prop}[defn]{Proposition}
\newtheorem{assum}[defn]{Assumption}
\newtheorem{ex}[defn]{Example}
\newtheorem{thm}[defn]{Theorem}
\newtheorem{cor}[defn]{Corollary}
\newtheorem{prob}{Problem}

\providecommand{\R}{\ensuremath \mathbb{R}}
\providecommand{\N}{\ensuremath \mathbb{N}}
\providecommand{\Z}{\ensuremath \mathbb{Z}}
\newcommand{\SE}{\regtext{SE}}
\newcommand{\SO}{\regtext{SO}}

\newcommand{\regtext}[1]{\mathrm{\textnormal{#1}}}
\newcommand{\ol}[1]{\overline{#1}}
\newcommand{\ts}[1]{\textsuperscript{#1}}
\newcommand{\lbl}[1]{_{\regtext{#1}}}
\newcommand{\etal}{\textit{et al}.}
\newcommand{\ie}{\textit{i}.\textit{e}.}
\newcommand{\eg}{\textit{e}.\textit{g}.}

\newcommand{\comp}{^{\regtext{C}}}
\newcommand{\card}[1]{\left\vert#1\right\vert}
\newcommand{\interior}[1]{\regtext{int}\!\left(#1\right)}
\newcommand{\proj}{\regtext{proj}}
\newcommand{\norm}[1]{\left\Vert#1\right\Vert}
\newcommand{\floor}[1]{\left\lfloor#1\right\rfloor}
\newcommand{\ceil}[1]{\left\lceil#1\right\rceil}
\newcommand{\abs}[1]{\left\vert#1\right\vert}
\newcommand{\pow}[1]{\regtext{pow}\!\left(#1\right)}
\newcommand{\diag}[1]{\regtext{diag}\!\left(#1\right)}
\newcommand{\eig}[1]{\regtext{eig}\!\left(#1\right)}
\newcommand{\union}{\bigcup}
\newcommand{\intersection}{\bigcap}
\newcommand{\trans}{^\intercal}
\newcommand{\inv}{^{-1}}
\newcommand{\pinv}{^{\dagger}}
\newcommand{\sign}{\regtext{sign}}
\newcommand{\expm}{\regtext{exp}}
\newcommand{\logm}{\regtext{log}}
\newcommand{\skw}{_{\times}}
\newcommand{\bigO}{\mc{O}}
\newcommand{\bdry}[1]{{\partial #1}}
\renewcommand{\ker}[1]{\regtext{ker}\!\left(#1\right)}
\newcommand{\convhull}[1]{\regtext{CH}\!\left(#1\right)}
\newcommand{\dist}[1]{d_2\!\left(#1\right)}
\newcommand{\len}[1]{\left\Vert#1\right\Vert_2}
\newcommand{\trace}{\regtext{trace}}
\newcommand{\degrees}[1]{$#1^\circ$}

\newcommand{\emptyarr}{[\ ]}
\newcommand{\zeros}{{0}}
\newcommand{\ones}{{1}}
\newcommand{\eye}{{I}}

\newcommand{\normaldist}{\mathcal{N}}
\newcommand{\est}[1]{\hat{#1}}

\newcommand{\state}{x}
\newcommand{\mean}{\mu}
\newcommand{\cov}{\Sigma}
\newcommand{\bearing}{\beta}
\newcommand{\eststate}{\est{\state}}
\newcommand{\mode}{y}
\newcommand{\ctrl}{u}
\newcommand{\noise}{w}
\newcommand{\error}{\varepsilon}
\newcommand{\position}{\mathbf{p}}
\newcommand{\course}{\psi}
\newcommand{\speed}{\varepsilon}
\newcommand{\surgespeed}{U}
\newcommand{\vel}{\mathbf{v}}
\newcommand{\probability}[1]{\Pr\left\lbrace#1\right\rbrace}
\newcommand{\condprob}[2]{\probability{#1 \; | \; #2}}
\newcommand{\set}[1]{\left\lbrace #1 \right\rbrace}
\newcommand{\condset}[2]{\set{#1 \; : \; #2}}

\renewcommand{\sin}[1]{\regtext{sin}\left( #1 \right)}
\renewcommand{\cos}[1]{\regtext{cos}\left( #1 \right)}
\renewcommand{\arctan}[1]{\regtext{arctan}\left( #1 \right)}

\newacronym{COLREGs}{COLREGs}{International Regulations for Preventing Collisions at Sea}
\newacronym{ECDIS}{ECDIS}{Electronic Chart Display and Information System}
\newacronym{CPA}{CPA}{Closest Point of Approach}
\newacronym{TCPA}{TCPA}{Time for Closest Point of Approach}
\newacronym{DCPA}{DCPA}{Distance at Closest Point of Approach}
\newacronym{AIS}{AIS}{Automatic Identification System}
\newacronym{GPS}{GPS}{Global Positioning System}
\newacronym{ENC}{ENC}{Electronic Navigational Chart}
\newacronym{INS}{INS}{Inertial Navigation System}
\newacronym{SOG}{SOG}{Speed Over Ground}
\newacronym{COG}{COG}{Course Over Ground}

\newacronym{DFA}{DFA}{Deterministic Finite-State Automata}
\newacronym{DES}{DES}{Discrete-Event Systems}
\newacronym{MPC}{MPC}{Model-Predictive Controller}
\newacronym{cvae}{CVAE}{Conditional Variational Autoencoder}
\newacronym{lstm}{LSTM}{Long Short-Term Memory}
\newacronym{cnn}{CNN}{Convolutional Neural Network}
\newacronym{gru}{GRU}{Gated Recurrent Unit}
\newacronym{RL}{RL}{Reinforcement Learning}
\newacronym{STL}{STL}{Signal Temporal Logic}

\newacronym{COLAV}{COLAV}{Collision Avoidance}
\newacronym{ROS}{ROS}{Rate of Swing}
\newacronym{TV}{TV}{Target Vessel}
\newacronym{OS}{OS}{Ownhip}
\newacronym{ASV}{ASV}{Autonomous Surface Vessel}
\newacronym{MASS}{MASS}{Marine Autonomous Surface System}
\newacronym{SIMAC}{SIMAC}{Svendborg International Maritime Academy}

\newacronym{KDE}{KDE}{Kernel Density Estimate}
\newacronym{ISJ}{ISJ}{Improved Sheather-Jones}

\newacronym{EKF}{EKF}{Extended Kalman Filter}

\newacronym{CR}{CR}{Collision Risk}
\newacronym{IMO}{IMO}{International Maritime Organization}
\newacronym{MC}{MC}{Monte Carlo}
\newacronym{ANN}{ANN}{Artificial Neural Network}
\newacronym{DNN}{DNN}{Deep Neural Network}
\newacronym{RNN}{RNN}{Recurrent Neural Network}
\newacronym{LSTM}{LSTM}{Long-Short Term Memory}
\newacronym{I/O}{I/O}{Input-Output}
\newacronym{ACC}{ACC}{Adaptive Cruise Control}
\newacronym{ACS}{ACS}{Autonomous Coordination Supervisor}
\newacronym{APS}{APS}{Autonomous Platform Supervisor}
\newacronym{ANS}{ANS}{Autonomous Navigation Supervisor}
\newacronym{SAS}{SAS}{Situation Awareness Service}
\newacronym{SA}{SA}{Situation Array}
\newacronym{SHP}{SHP}{Short Horizon Planner}
\newacronym{COL}{COL}{Consolidated Object List}
\newacronym{FSA}{FSA}{Finite-State Automata}
\newacronym{RCC}{RCC}{Remote Control Center}

\title{Stochastic COLREGs Evaluation for Safe Navigation under Uncertainty}

\author{
    Peter Nicholas Hansen\ts{1},
    Dimitrios Papageorgiou\ts{1},
    Roberto Galeazzi\ts{1},
    Mogens Blanke\ts{1}
\thanks{\ts{1}Automation and Control Group, Department of Electrical and Photonics Engineering, Technical University of Denmark, Elektrovej 326, 2800 Kgs Lyngby, Denmark}
\thanks{This research was part of the ShippingLab Autonomy work package, sponsored by Innovation Fund Denmark, the Danish Maritime Fund, Orients Fund and Lautitzen Fonden under grant number 8090-00063B.
    Corresponding author: Peter Nicholas Hansen, \texttt{pnha@dtu.dk}.}
}

\maketitle

\begin{abstract}
    The encounter situation between marine vessels determines how they should navigate to obey COLREGs, but time-varying and stochastic uncertainty in estimation of angles of encounter, and of closest point of approach, easily give rise to different assessment of situation at two approaching vessels. This may lead to high-risk conditions and could cause collision. This article considers decision making under uncertainty and suggests a novel method for probabilistic interpretation of vessel encounters that is explainable and provides a measure of uncertainty in the evaluation. The method is equally useful for decision support on a manned bridge as on Marine Autonomous Surface Ships (MASS) where it provides input for automated navigation. The method makes formal safety assessment and validation feasible.  We obtain a resilient algorithm for machine interpretation of COLREGs under uncertainty and show its efficacy by simulations.
\end{abstract}


\section{Introduction}\label{sec:introduction}

\IEEEPARstart{W}{hen} a navigator on watch anticipates a  risk of collision with another vessel at sea, he/she will assess the situation and, based on the information available, act to navigate safely. 
Uncertainty in interpretation of angles of encounter, and of the intentions of other vessels, are prime causes for mutual misinterpretations and risk of collision.
The topic for this research is hence to establish situational awareness under uncertainty, and develop computer algorithms for decision-making under uncertainty.

Situational awareness determines whether one or more targets pose a risk. \emph{Situational awareness} has three distinct steps in cognitive sciences \cite{Endsley1995}: 1) \emph{perception}, 2) \emph{understanding} \& 3) \emph{anticipation}. 
Fusing all available information from navigation sensors and outlook, \emph{perception} establish a coherent representation of the surroundings.
\emph{Understanding} attaches a higher semantic meaning to the information and determines which navigation rules apply.
\emph{Anticipation} predicts how the future is believed to evolve and whether a situation is likely to pose a risk. 
\emph{understanding} and \emph{anticipation} are iterated if this is needed.
 Marine vessels are obliged to navigate according to \gls{COLREGs} if a risk is apparent. Uncertainty in sensor information, and on interpretation of other vessel's intentions, cause uncertainty on the overall assessment of obligations, and mitigation of uncertainty driven risks are essential, but have not been in focus in earlier research. This paper therefore employs stochastic methods in a novel algorithm, offering safe navigation decisions under uncertainty.

The effect of estimation uncertainty has been examined in different fields of autonomous systems, in particular for self-driving cars and unmanned aerial vehicles, but also for the maritime vessels.
\gls{DNN} based methods are widely used in highly automated cars. Early end-to-end implementations had a single model estimating a control action based on raw sensory input \cite{Bojarski2016} but better results were achieved with divided architectures, \cite{Chai2019, Zhao2020, Tolstaya2021} where dedicated \gls{DNN} models predict trajectories of individual agents in the surroundings, e.g. vehicles and pedestrians.
Including a dynamical model \cite{Salzmann2020}, to predict accelerations of nearby agents, the physical model ensured that generated trajectories were dynamically feasible.
Common for these \gls{DNN} based methods are that uncertainty was generated by sampling the estimated embedding distributions of the model.

The use of \gls{DNN} was also examined in a maritime context.
Data-sets collected from an inland container vessel, with \gls{AIS}, radar and \gls{ENC} recordings, were used  by \cite{Dijt2020} to train a \gls{DNN} to predict future trajectories of other inland vessels.
Such use of historical data and a \gls{DNN} were able to accurately predict the future trajectories of marine vessels \cite{Murray2018, Murray2019, Schoeller2021}, representing the fact that there are preferred corridors and routes preferred by marine vesels in near-coast operatrion.
\gls{DNN} based methods for probabilistic prediction of vessels based on \gls{AIS} data were further explored in \cite{Schoeller2021a, Soerensen2022}.
Common for all these maritime prediction models, is  that models predict only a \emph{marginal probability} for trajectory of a target, they do not account for nearby vessels or other objects.

Methods to asses probabilistic properties of predictions were dealt with in \cite{Tiger2020}, who extended \gls{STL} to reason about the uncertainty of their own belief e.g., the probability that a minimum or maximum height threshold was violated for unmanned aerial vehicles.
\gls{STL} was also used to formulate and evaluate \gls{COLREGs} obligations and compliance of marine vessels \cite{Krasowski2021}, however, in a non-probabilistic framework.

The use of \gls{COLREGs} was researched extensively in the field of \gls{COLAV}, where  \cite{Burmeister2021} employed several methods.
Further, \gls{RL} algorithms were investigated to learn a \gls{COLREGs}-compliant control policy for \gls{COLAV}, combining simulations of specific scenarios \cite{Zhao2019}, random ones \cite{Meyer2020a}, and randomly generated from historical \gls{AIS} data \cite{Meyer2020}.
Implementation of ``expert knowledge" using a multi-objective particle swarm approach \cite{Hu2019}, showed the ability for a \gls{COLAV} algorithm to correctly favour a ``good seamanship" approach in compliance with \gls{COLREGs}, by letting the algorithm favour course-change over speed-change in giving-way scenarios.
\gls{MPC} based methods was also explored for \gls{COLAV}. It showed robust performance, and maintained \gls{COLREGs} compliance \cite{Thyri2022}.
A Branching-Course Model-Predictive Controller (BC-MPC) was specifically designed \cite{Eriksen2019} to ensure robustness towards target prediction uncertainty in a path-planning problem.
Inclusion in \gls{MPC} of uncertainty for a target obstacle was examined in \cite{Johansen2016, Kufoalor2019}. They accounted for uncertainty by creating a span of possible trajectories for a target by adding and subtracting arbitrarily chosen $\delta$ to the estimated course and speed of the target.
The proposed method was demonstrated to be able to handle both single and multi-vessel scenarios while maintaining \gls{COLREGs} compliance.
However, the uncertainty representation does not reflect the actual uncertainty of the prediction model, which may be either to conservative or not conservative enough, depending on the situation.

Fuzzy logic methods were early suggested for use in marine navigation. 
Algorithms included \cite{Hasegawa1987} for collision-avoidance, \cite{Hasegawa2012} for automated simulator manoeuvring of multiple target vessels, and \cite{Perera2012} for general \gls{COLAV}. 
The latter generated a sequence of actions, enabling  collision avoidance in a multi-vessel encounter.  
\cite{Bakdi2022} extended to accounted for more than the commonly used rules 13-15 in the perhaps most \gls{COLREGs}-complete decision-system based on fuzzy logic.
While fuzzy logic implementations naturally incorporate a notion of uncertainty in their design, and the decisions/outputs of the models are ``explainable", unlike some of the learning based methods, the method requires tuning of parameters. 
It was argued that ``expert knowledge" can be used for the tuning of these parameters \cite{Bakdi2022}. 
While the value of each parameter might be simple to explain, the interplay between them is far from trivial when the entire envelope of different scenarios need be considered.

A probabilistic approach to include risk assessment in decision-making for \gls{COLAV} was explored in \cite{Tengesdal2020a}.
An extension in \cite{Tengesdal2020} modelled targets' intention by selecting either of three actions: keep course and speed; turn starboard; turn port. 
This required a-priori knowledge about target's way point(s) and intention to follow \gls{COLREGs}.
The variance of the sampled risk uncertainty was reduced in \cite{Tengesdal2022}, using importance sampling. 
Despite being comprehensive in most aspects, these results were restricted to the case of \gls{OS} being one of the two vessels involved in an encounter.
A generic situation awareness algorithm should be able to let \gls{TV}s interact with each other as needed to comply to their mutual obligations.

The present paper considers generic situation awareness where \gls{TV}s manoeuvre according to other vessels in their surroundings and not only according to \gls{OS} - \gls{TV} obligations. This will be shown to provide a framework to handle decision support under uncertainty in scenarios with multiple gls{TV}s interacting both mutually and with \gls{OS}. The approach is a probabilistic interpretation of collision risk assessment, building on \gls{TCPA},\gls{DCPA} calculations.
The method uses  Monte-Carlo methods of sampling, based on estimation of uncertainty for \gls{TV}s, enabling the system to make \gls{COLREGs}-compliant decisions in uncertain situations, which increases the the overall compliance with \gls{COLREGs} regulations.
The contribution of this work include: 1) the probabilistic evaluation of risk of collision (Rule 7(a)); 2) the probabilistic interpretation of a given situation (Rule 13-14(c)). 
Both contributions ensures that \gls{COLREGs} compliance is achieved.
The proposed methods will be shown to apply on any pair of vessels, and hence provides genuine situation awareness that anticipate the manoeuvres of \gls{TV}s in the entire area of attention.
The paper shows how this could be incorporated in an existing \gls{DES}-based framework for situational awareness, \cite{Hansen2020, Hansen2022, Papageorgiou2019, Papageorgiou2022},  and illustrates how the method can increase the safety of an autonomously navigating surface vessel.
The paper shows that the tracking problem is non-stationary, and highly-nonlinear due to transformations with origin in the calculation of \gls{TCPA}/\gls{DCPA}. The paper discusses how a sampling based approach solves thes complexities.
The paper furthermore presnt the benefits of implementing the uncertainty propagation into a \gls{DES} based framework, and compares with a purely distribution based approach.

The paper is structured as follows; section \ref{sec:background} provides an overview of the tools and methods used in maritime navigation.
Section \ref{sec:method} gives a detailed analysis and description of how the methods shown in section \ref{sec:background} can be utilized in a probabilistic toolset, and illustrates how the probabilistic tools is included into the already existing \gls{COLREGs}-compliant framework.
In section \ref{sec:experiements} three different simulation scenarios are illustrated; one scenario is for the analysis of the effect of uncertainty in the decision-making, and two scenarios highlight how the automata-based implementation enables a safer decision-making system.
Appendix \ref{app:COLREGS} provides a brief explanation of the relevant \gls{COLREGs} rules.
Finally, section \ref{sec:discussion} discuss the proposed method and results,  and \ref{sec:conclusion} provides a conclusion and future research avenues.

\section{Motivation}

In case of uncertainty of the type of situation, specifically in head-on and overtaking situations, and thereby uncertainty of the responsibilities of ownship, a human navigator must assume that ownship has the responsibility to give-way, as stated in rules 13 \& 14 \cite{IMO2021, Cockcroft2011}.
\begin{figure}[!htb]
	\centering
	\tikzsetnextfilename{port_headon}
		\resizebox{0.6\columnwidth}{!}{\includegraphics{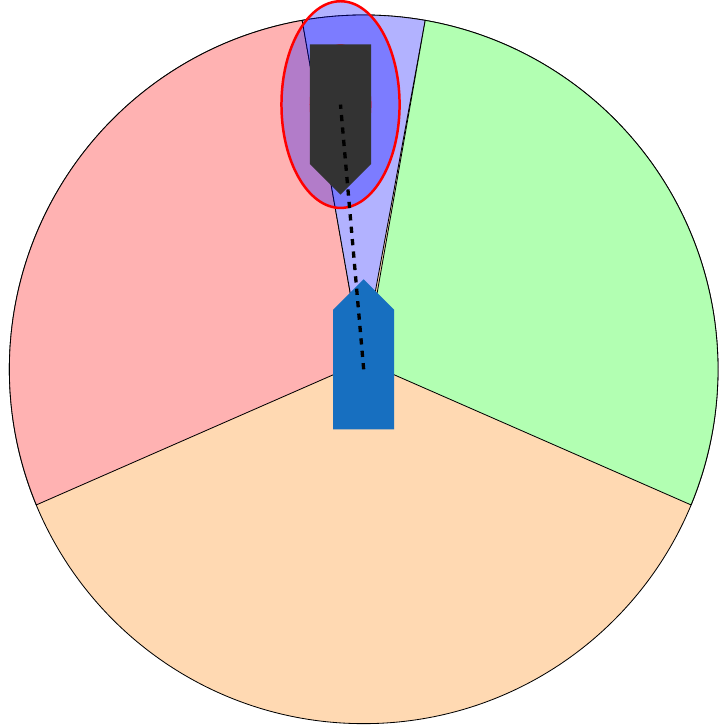}}
	\caption{An ambiguous situtation where the target vessel (black dot) with uncertainty ellipse (shown in red) can result in the situation being interpreted as; 1) in a head-on (\textcolor{RoyalBlue!70}{blue}), 2) a port-side crossing (\textcolor{Red!70}{red}) or 3) a starboard crossing (\textcolor{Green!70}{green}).}
	\label{fig:port_headon}
\end{figure}
The effect of this uncertainty in the situation type, is highest in the border between the situations where the responsbilities of ownship changes, i.e., the border between a port-side crossing (rule 15) and a head-on (rule 14), as this inverts the responsbilities of ownship from ``stand-on'' to ``give-way''.
\figref{fig:port_headon} illustrates the previously given example situation, where the target vessel (illustrated as the black circle), is on the border between the head-on and port-crossing regions.
In this situation the human navigator will looking for the visual indicators of the orientation of the target vessel, e.g., are both the port and starbord lanterns visible, indicating that the target vessel is on a course that is \degrees{180}$\pm$ \degrees{5} opposite of ownship course, indicating that the situation is a ``head-on'' encounter.
However, as vessels can have differing heading and \gls{COG} due to wind and current, visual indication of the aspect might not suffice for correct interpretation of the situation.

The above necessitates the need for an autonomous sytem to be able to reason about the uncertainty of the current situation around itself, if it is to safely navigate in an environment alongside humans.
As an example \figref{fig:port_headon} clearly illustrates why a deterministic interpretation of the situation could lead to a clearly incorrect assessment for an autonomous system, i.e., if a sample was to be drawn from the positional distribution of the target vessel (illustrated as the blue ellipses with a red border), this could result in a ``starboard-crossing'' evaluation of the situation, which is incorrect, or in worst-case, be evaluated as a ``port crossing".

\section{Background}\label{sec:background}

The comfort zone is a boundary that is defined by the navigator, and is determined by the manoeuvrability of the vessel, and by the surroundings. The comfort zone in open sea is much larger than that in confined waters.

Calculation of \gls{CPA} of two vessels requires data for both vessels: \emph{position}, \emph{course} and \emph{speed}.
Let 
\begin{equation}
    \position_{j} = \left[N_{j},E_{j}\right]\trans, \quad
    \position_{k} = \left[N_{k},E_{k}\right]\trans,
\end{equation}
be the position of vessel $j$ and $k$ in a North-East tangent plane.  Tangent plane velocities are
\begin{align}
    \vel_{j} &= \left[\surgespeed_{j}\sin{\course_{j}},\surgespeed_{j}\cos{\course_{j}}\right]\trans,    \\
    \vel_{k} &= \left[\surgespeed_{k}\sin{\course_{k}},\surgespeed_{k}\cos{\course_{k}}\right]\trans,
\end{align}
be the velocity vectors for vessels $j$ and $k$, respectively, in the same frame, where $\surgespeed_{(\cdot)}$ is the vessels' speed and $\course_{(\cdot)}$ it's course. 
With the position, forward velocities and course of both vessels, the \gls{TCPA} is defined as
\begin{align}\label{eq:tcpa}
    \mathrm{TCPA} &\triangleq - \frac{\left(\position_{j} - \position_{k}\right)\trans \left(\vel_{j} - \vel_{k}\right)}{\left(\vel_{j} - \vel_{k}\right)\trans \left(\vel_{j} - \vel_{k}\right)} \nonumber \\
    &=- \frac{\left(\Delta \position\right)\trans \Delta \vel}{\left(\Delta \vel\right)\trans \Delta \vel}.
\end{align}
If both vessels maintain speed, their respective positions at time TCPA are predicted as
\begin{align}
    \mathbf{p}_{j}(\mathrm{TCPA}) &= \mathbf{p}_{j} + \mathbf{v}_{j} \cdot \mathrm{TCPA}, \\
    \mathbf{p}_{k}(\mathrm{TCPA}) &= \mathbf{p}_{k} + \mathbf{v}_{k} \cdot \mathrm{TCPA}.
\end{align}
The \gls{DCPA} is then defined as follows
\begin{equation}\label{eq:dcpa}
    \mathrm{DCPA} \triangleq \norm{\Delta \mathbf{p}(\mathrm{TCPA})}.
\end{equation}
Finally, let $\bearing_{j}^{k}$ be the relative bearing to vessel $k$ from vessel $j$, defined as
\begin{equation}\label{eq:bearing_deterministic}
    \bearing_{j}^{k} \triangleq \frac{180}{\pi} \arctan{\frac{ E_{k} - E_{j} }{N_{k} - N_{j}}} - \course_{j}.
\end{equation}

\subsection{Behavioural Model}

\begin{figure}[tb]
	\centering
	\tikzsetnextfilename{colregs_regions}
		\resizebox{0.8\columnwidth}{!}{\includegraphics{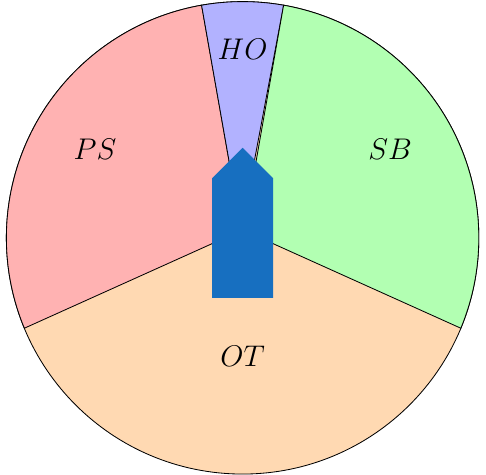}}
	\caption{A visualisation of how $g_1$ maps a bearing $\bearing$ into the different regions: head-on (HO), starboard (SB), overtaking (OT) and port (PS)}
	\label{fig:colregs_regions}
\end{figure}

The behavioral model of a vessel concerns its obligation to give-way or stand-on given a specific situation.
Upon an encounter with another vessel, two elements need be determined; 1) the existence of a collision risk, 2) the applicable \gls{COLREGs} rule.
The first relates to comparing \gls{DCPA} and \gls{TCPA} values to given thresholds, i.e. 
\begin{align}
    \mathrm{DCPA} &\leq d_{act} \\
    \mathrm{TCPA} &\leq t_{aware}.
\end{align}
The second requires associating the spatial configuration of the two vessels to one of four possible regions, shown in Fig.~\ref{fig:colregs_regions}.

If a risk of collision is deemed to exist, the configuration of the two vessels is taken into account when applying the \gls{COLREGs} rules e.g., port or starboard crossing, etc.
The configuration is determined from the relative bearing $\bearing$ between the two vessels.
\begin{definition}
    \textbf{(Bearing Mapping)}:
    Let $g_1: \mathcal{B}^3 \rightarrow \mathcal{S}_1$ be a function that maps the courses $\course_j, \course_k$ and relative bearing $\bearing \in \mathcal{B} \triangleq \condset{\bearing \in \R}{0 \leq \bearing < 360}$ into one of the four regions $\mathcal{S}_1 \triangleq \set{HO, SB, OT, PS}$, according to
    \begin{align}\label{eq:bearing_zone_map}
    	&g_1(\bearing, \course_j, \course_k) = \nonumber \\
        &\begin{cases}
        	HO \text{, for } \left( 0 \leq \bearing \leq 5 \right) \lor \left( 355 < \bearing < 360 \right) \\ 
         \qquad \qquad \lor  \left( \abs{\Delta \course} \leq 5 \right) \\  %
        	SB\text{, for } \left( 5 < \bearing \leq 112.5 \right) \land  \left(  \abs{\Delta \course} > 5 \right) \\
        	OT\text{, for } \left( 112.5 < \bearing \leq 247.5  \right) \land  \left(  \Delta \course > 5 \right)  \\
        	PS\text{, for } \left( 247.5 < \bearing \leq 355  \right) \land  \left(  \Delta \course > 5 \right) ,
        \end{cases}
    \end{align}
    where $HO$, $SB$, $OT$ \& $PS$ are the ``Head-on", ``Starboard", ``Overtaking" and ``Port" regions, respectively, and where the reciprocal course $\Delta \course \in [-180, 180]$ is defined as
    \begin{equation}\label{eq:headon_course_mapping}
    	\Delta \course = \mod \left( \left( \course_j - \course_k \right), 360 \right) - 180, 
    \end{equation}
    such that two reciprocal \gls{COG}'s result in $\Delta \course = 0$.
\end{definition}
In order to evaluate a situation and apply the correct \gls{COLREGs} rule, the relative bearing must be evaluated from the perspective of both vessels.

In order to evaluate the COLREGs situation, the relative bearing must be evaluated from both vessels, this makes the following definition useful.

Consider the encounter of two arbitrary vessels $j$ and $k$, where each gives the following region mapping of the other vessel
\begin{align}
	v_{j} &= g_1(\bearing_{j}^{k}, \course_{j}, \course_{k})   \\
	v_{k} &= g_1(\bearing_{k}^{j}, \course_{k}, \course_{j})
\end{align}
and $v_{j}, v_{k} \in \mathcal{S}_1$.

\begin{rem}\normalfont
The situation can be interpreted from the viewpoint of both vessel $j$ and $k$, the term \emph{acting agent} will be used to refer to the vessel that assess the sistuation to make a decision.
The term \emph{target agent} will be reserved for any other vessel in the scene.
In essence, acting agent refers to the vessel under which perspective, the situation is assessed.

\end{rem}
\begin{definition}
    \textbf{(Mutual Bearing Mapping)}: \\
    Define the mapping $g_2: \mathcal{S}_1 \times \mathcal{S}_1 \rightarrow \mathcal{S}_2$, where
    \begin{equation}\label{eq:bearing_colregs_map}
        \mathcal{S}_2 \triangleq \condset{(p, q)}{p \in \set{R_{0}, R_{13}, R_{14}, R_{15}}, q \in \{0, 1\}},
    \end{equation}
    such that the elements in $S_2$ are tuples, with the first element being the set of \gls{COLREGs} rules, $R_{n}$, and $n$ being the  rule number. 
    The second element is the give-way/stand-on obligation (1/0 respectively) of the acting agent.
\end{definition}
For example, vessel $j$ would not be able to distinguish between an overtaking (rule 13) and a head-on (rule 14) situation with vessel $k$ placed at bearing $\bearing_j^k = 0^\circ$, without considering its own relative bearing towards vessel $k$, i.e., evaluating the situation from the point of view of vessel $k$.

Table~\ref{tab:colregs_mapping} describes the mapping $g_2$.
\begin{table}[tb]
    \centering
    \caption{Table illustrates how the function $g_2$ maps from two regions $v_{A}$ (acting agent) and $v_{T}$ (target agent) into the elements of $\mathcal{S}_2$. Each element in $\mathcal{S}_2$ contains two sub-elements; the first representing the applicable \gls{COLREGs} rule and the second represents the obligation of the vessel that is mapped to $v_1$, where $1=$``give-way", and $0=$``stand-on". Elements marked in red are situations where rules 13-15 cannot be applied.}
    \label{tab:colregs_mapping}
    \begin{tabular}{c|c c c c}
        \toprule
        \diagbox{$v_{A}$}{$v_{T}$} & $HO$ & $SB$ & $OT$ & $PS$ \\
        \midrule
        $HO$ & $(R_{14}, 1)$ & $(R_{15}, 0)$ & $(R_{13}, 1)$ & $(R_{15}, 1)$ \\
        $SB$ & $(R_{15}, 1)$ & \textcolor{Red}{$(R_0, 1)$} & $(R_{13}, 1)$ & $(R_{15}, 1)$ \\
        $OT$ & $(R_{13}, 0)$ & $(R_{13}, 0)$ & \textcolor{Red}{$(R_0, 1)$} & $(R_{13}, 0)$ \\
        $PS$ & $(R_{15}, 0)$ & $(R_{15}, 0)$ & $(R_{13}, 1)$ & \textcolor{Red}{$(R_0, 1)$} \\
        \bottomrule
    \end{tabular}
\end{table}
Note that the products $\set{(SB,SB), (OT, OT), (PS, PS)}$ all map to the element $(R_0, 1)$, with $R_0$ denoting that none of the \gls{COLREGs} rules 13-15 can be applied.
This is partly because these combinations are ``artificial constructs", meaning that these situations can only be created by either transitioning from one of the other situations, in such a case, the risk will/must have been mitigated according to the previous situation, or when turning the system on in the middle of a manoeuvre/situation.
However for completeness these situations are mapped to a more conservative ``give-way" obligation.
\begin{rem}\normalfont
    The function $g_2$ assumes that both vessels are power-driven.
    If either of the vessels are not power-driven, it changes the function mapping.
\end{rem}

\section{Method}\label{sec:method}

In the proposed framework, it is assumed that a sensor fusion module provides state estimates of  surrounding objects.
Each tracked object (target) $i$ will have a state vector $\state_i$ defined as,

\begin{equation}
    \state_i = [N_i, E_i, \course_i, \surgespeed_i]\trans,
\end{equation}
where $N_i$ and $E_i$ are the north and east coordinates, $\course_i$ is the \gls{COG}, and $\surgespeed_i$ is the velocity of the $i$-th  target.
The sensor fusion module utilizes an \gls{EKF} which is able to provide an estimate of the estimation error. 
As such, the sensor fusion module provides the state estimate $\eststate_i$, which is defined as the true state $\state_i$ plus some estimation error $\error_i$
\begin{equation}
    \eststate_i = \state_i + \error_i,
\end{equation}
where $\error_i$ is a Gaussian distribution with 
\begin{equation}\label{eq:estimation_error_def}
    \error_i \sim \normaldist(0, \cov_i).
\end{equation}
\begin{rem}\normalfont
    While sensor noise is ``low" compared to the effect of wave disturbances in the tracking of a target, and wave disturbances are most certaintly not guassian.
    The authors assume that the sensor fusion module contains a wave-filter, that will filter out the contribution of uncertainty that stems from wave disturbances in the sensor noise, which ensures that the previously mentioned assumption of guassianity of the estimation error is valid.
\end{rem}
For an overview of the architecture of the implemented system the reader is referred to \cite{Dittmann2021}, for details regarding the sensor fusion please see \cite{Dagdilelis2022a} or \cite{Dagdilelis2022}.

\subsection{Stochastic Behaviour Model}

By introducing stochasticity from the tracking error, the evaluation of the \gls{TCPA}, \gls{DCPA} values, and the mapping function $g_2$, must be done in a probabilistic context, i.e. by integration over specific density functions.
As such, the probability that a risk of collision exists between two vessels $j$ and $k$, with the estimated states $\eststate_j$ and $\eststate_k$ respectively, as assessed from the perspective of vessel $j$, i.e., vessel $j$ is the acting agent, is
\begin{equation}\label{eq:prob_dcpa}
    \condprob{DCPA \leq d_{act}}{\eststate_{j}, \eststate_{k}} = \int_{0}^{d_{act}} f_{DCPA}(x) \,dx,
\end{equation}
where $f_{DCPA}$ is the \gls{DCPA} density function.
It should be noted that since $DCPA$ can only assume non-negative values, the evaluation of the integral in \eqref{eq:prob_dcpa} is done from 0 instead of $-\infty$.
For brevity, the explicit condition on $\eststate_{j}$ and $\eststate_{k}$ in the notation $\probability{\cdot | \eststate_{j}, \eststate_{k}}$ is omitted for the remainder of the paper.
Similarly, the probability that the collision risk will occur within a certain time interval $[0,t_{aware}]$ is %
\begin{equation}\label{eq:prob_tcpa}
    \probability{0 \leq TCPA \leq t_{aware}} = \int_{0}^{t_{aware}} f_{TCPA}(x) \,dx ,
\end{equation}
where $f_{TCPA}$ is the \gls{TCPA} density function.
Since only future events are of interest, the lower limit is zero, despite \gls{TCPA} values belonging to the entirety of $\R$.
Given a risk of collision, the probability of a specific situation $\probability{s = S}$ where $S \in \mathcal{S}_2$
is
\begin{equation}\label{eq:prob_situation}
    \probability{s = S} = \probability{v_1 = A, v_2=B},
\end{equation}
where $A, B \in \mathcal{S}_1$ are the bearing mappings from each vessel's perspective.
Since the two mappings are not correlated (the heading of one vessel does not impact the heading of the other vessel), \eqref{eq:prob_situation} can be rewritten as
\begin{equation}
    \probability{v_1 = A, v_2 = B} = \probability{v_1 = A} \probability{v_2 = B}.
\end{equation}
The individual probabilities of a relative bearing measurement  belonging to each of the regions in Fig.~\ref{fig:colregs_regions} (specific element $v$ of $\mathcal{S}_1$) are
\begin{align}
    \probability{v = HO} &= \int_{0}^{5} f_{\bearing}(\bearing) \,d\bearing + \int_{355}^{360} f_{\bearing}(\bearing) \,d\bearing \\
   \probability{v = SB} &= \int_{5}^{112.5} f_{\bearing}(\bearing) \,d\bearing \\
   \probability{v = OT} &= \int_{112.5}^{247.5} f_{\bearing}(\bearing) \,d\bearing \\
   \probability{v = PS} &= \int_{247.5}^{355} f_{\bearing}(\bearing) \,d\bearing.
\end{align}
where $f_{\bearing}$ is the bearing density function.
From \eqref{eq:prob_situation} and \eqref{eq:bearing_colregs_map} it is possible to calculate the probability that vessel $j$ needs to give way, i.e. as the sum of probabilities of all the elements of $G \subset \mathcal{S}_2$
\begin{align}\label{eq:give_way_set}
    G = \{  &(HO, HO), (HO, OT), (HO, PS), \nonumber \\ 
            &(SB, HO), (SB, SB), (SB, OT), \nonumber \\ 
            &(SB, OT), (OT, OT), (PS, OT), (PS, PS) \},
\end{align}
where $G$ contains all the elements of $\mathcal{S}_2$ where the second element $q$ in the tuple $(p,q)$ is $q=1$.
Therefore, given a risk of collision, the probability that vessel $j$ needs to give-way is
\begin{equation}\label{eq:give_way_prob}
    \condprob{\text{``give-way"}}{DCPA \leq d_{act}} = \sum_{g\in G} \probability{s=g}.
\end{equation}
and the total probability of vessel $j$ being required to give way equals
\begin{align}
    \probability{\text{``give-way"}} =& \condprob{\text{``give-way"}}{DCPA \leq d_{act}}\nonumber\\
    & \cdot \probability{DCPA \leq d_{act}} \nonumber \\
    =& \sum_{g\in G} \probability{s=g} \cdot \probability{DCPA \leq d_{act}}. \label{eq:give_way_prob_total}
\end{align}
From \eqref{eq:give_way_prob_total} it follows that the probability that vessel $j$ should stand on is
\begin{equation} \label{eq:stand_on_prob_total}
    \probability{\text{``stand-on"}} = 1 - \sum_{g\in G} \probability{s=g} \cdot \probability{DCPA \leq d_{act}}.
\end{equation}

\subsection{Analysis of Uncertainty Propagation}

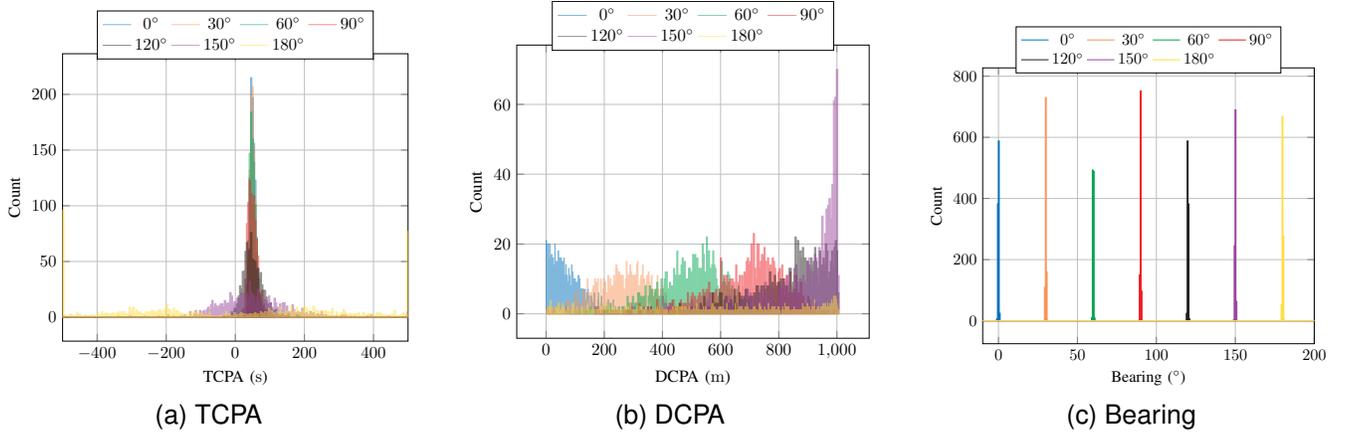
\begin{figure*}[tb]
    \centering
    \tikzsetnextfilename{tcpa_hist}
    \subfloat[TCPA\label{fig:tcpa_hist}]{
        \resizebox{0.31\textwidth}{!}{
            \begin{tikzpicture}
    \def\nbins{50}
	\def\opacity{0.5}
    \begin{axis} [
        width=\columnwidth,
        scaled ticks=false,
        xmin=-500,
        xmax=500,
        legend style={at={(0.5,1.15)}, anchor=north,legend columns=4},     
        grid,
        ylabel={Count},
        xlabel={TCPA (\unit{\second})}
    ]
    \addplot[ybar,fill,width=1pt,color=RoyalBlue, opacity=\opacity, hist={bins=2\nbins}] table [y=tcpa, col sep=comma] {data/bearing_0_data.csv};
    \addlegendentry{\degrees{0}}
    
    \addplot[ybar,fill,width=1pt,color=Peach, opacity=\opacity, hist={bins=2\nbins}] table [y=tcpa, col sep=comma] {data/bearing_30_data.csv};
    \addlegendentry{\degrees{30}}
    
    \addplot[ybar,fill,width=1pt,color=Green, opacity=\opacity, hist={bins=2\nbins}] table [y=tcpa, col sep=comma] {data/bearing_60_data.csv};
    \addlegendentry{\degrees{60}}
    
    \addplot[ybar,fill,width=1pt,color=Red, opacity=\opacity, hist={bins=2\nbins}] table [y=tcpa, col sep=comma] {data/bearing_90_data.csv};
    \addlegendentry{\degrees{90}}

    \addplot[ybar,fill,width=1pt,color=Black, opacity=\opacity, hist={bins=2\nbins}] table [y=tcpa, col sep=comma] {data/bearing_120_data.csv};
    \addlegendentry{\degrees{120}}

    \addplot[ybar,fill,width=1pt,color=Purple, opacity=\opacity, hist={bins=2\nbins}] table [y=tcpa, col sep=comma] {data/bearing_150_data.csv};
    \addlegendentry{\degrees{150}}

    \addplot[ybar,fill,width=1pt,color=Goldenrod, opacity=\opacity, hist={bins=2\nbins}] table [y=tcpa, col sep=comma] {data/bearing_180_data.csv};
    \addlegendentry{\degrees{180}}
    
    \end{axis}
\end{tikzpicture}
        }
    }
    \hfill
    \tikzsetnextfilename{dcpa_hist}
    \subfloat[DCPA\label{fig:dcpa_hist}]{
        \resizebox{0.31\textwidth}{!}{
            \begin{tikzpicture}
    \def\nbins{200}
	\def\opacity{0.5}
	\begin{axis} [
		width=\columnwidth,
		scaled ticks=false,
		legend style={at={(0.5,1.15)}, anchor=north,legend columns=4},
		grid,
		ylabel={Count},
		xlabel={DCPA (\unit{\m})}
		]
	
		\addplot[ybar,fill,color=RoyalBlue, opacity=\opacity, hist={bins=\nbins}] table [y=d_cpa, col sep=comma] {data/bearing_0_data.csv};
		\addlegendentry{\degrees{0}}
  		
		\addplot[ybar,fill,color=Peach, opacity=\opacity, hist={bins=\nbins}] table [y=d_cpa, col sep=comma] {data/bearing_30_data.csv};
		\addlegendentry{\degrees{30}}
	
        \addplot[ybar,fill,color=Green, opacity=\opacity, hist={bins=\nbins}] table [y=d_cpa, col sep=comma] {data/bearing_60_data.csv};
		\addlegendentry{\degrees{60}}
		
		\addplot[ybar,fill,color=Red, opacity=\opacity, hist={bins=\nbins}] table [y=d_cpa, col sep=comma] {data/bearing_90_data.csv};
		\addlegendentry{\degrees{90}}
		
		\addplot[ybar,fill,color=Black, opacity=\opacity, hist={bins=\nbins}] table [y=d_cpa, col sep=comma] {data/bearing_120_data.csv};
		\addlegendentry{\degrees{120}}
	
		\addplot[ybar,fill,color=Purple, opacity=\opacity, hist={bins=\nbins}] table [y=d_cpa, col sep=comma] {data/bearing_150_data.csv};
		\addlegendentry{\degrees{150}}

		\addplot[ybar,fill,color=Goldenrod, opacity=\opacity, hist={bins=2\nbins}] table [y=d_cpa, col sep=comma] {data/bearing_180_data.csv};
		\addlegendentry{\degrees{180}}		
	\end{axis}
\end{tikzpicture}
        }
    }
    \hfill
    \tikzsetnextfilename{bearing_hist}
    \subfloat[Bearing\label{fig:bearing_hist}]{
        \resizebox{0.31\textwidth}{!}{
            \begin{tikzpicture}
    \begin{axis} [
        width=\columnwidth,
        scaled ticks=false,
        xmin=-10,
        xmax=200,
        legend style={at={(0.5,1.15)}, anchor=north,legend columns=4},     
        grid,
        ylabel={Count},
        xlabel={Bearing ($^\circ$)}
    ]
    \addplot[ybar,fill,width=1pt,color=RoyalBlue, hist={bins=360}] table [y=bearing, col sep=comma] {data/bearing_0_data.csv};
    \addlegendentry{\degrees{0}}
    
    \addplot[ybar,fill,width=1pt,color=Peach, hist={bins=360}] table [y=bearing, col sep=comma] {data/bearing_30_data.csv};
    \addlegendentry{\degrees{30}}
    
    \addplot[ybar,fill,width=1pt,color=Green, hist={bins=360}] table [y=bearing, col sep=comma] {data/bearing_60_data.csv};
    \addlegendentry{\degrees{60}}
    
    \addplot[ybar,fill,width=1pt,color=Red, hist={bins=360}] table [y=bearing, col sep=comma] {data/bearing_90_data.csv};
    \addlegendentry{\degrees{90}}

    \addplot[ybar,fill,width=1pt,color=Black, hist={bins=360}] table [y=bearing, col sep=comma] {data/bearing_120_data.csv};
    \addlegendentry{\degrees{120}}

    \addplot[ybar,fill,width=1pt,color=Purple, hist={bins=360}] table [y=bearing, col sep=comma] {data/bearing_150_data.csv};
    \addlegendentry{\degrees{150}}

    \addplot[ybar,fill,width=1pt,color=Goldenrod, hist={bins=360}] table [y=bearing, col sep=comma] {data/bearing_180_data.csv};
    \addlegendentry{\degrees{180}}
    
    \end{axis}
\end{tikzpicture}    
        }
    }
    \caption{Histograms of the Calculated \gls{TCPA} and \gls{DCPA} values for the 7 different scenarios. Note that the \gls{TCPA} values are truncated in the histograms, such that the count in the lowest and highest bins are ``artificially" high. }
    \label{fig:histograms}
\end{figure*}

From \eqref{eq:prob_dcpa}-\eqref{eq:prob_situation} it is evident that the density functions $f_{TCPA}$, $f_{DCPA}$ and $f_{\bearing}$ are needed for a probabilistic evaluation of a given situation.
Given that the state variables in the estimated state of a tracked target $i$ is provided with a specific uncertainty $\cov_i$ as stated in \eqref{eq:estimation_error_def}, it is natural to ask the question ``how does this uncertainty propagate through the \gls{TCPA}, \gls{DCPA} and bearing equations?".
To answer that question, let $X_i$, $Y_i$, $\Psi_i$ and $V_i$ be random variables of the North, East, course and forward velocity of the $i$-th vessel, such that
\begin{equation}\label{eq:state_stochastic}
    \begin{bmatrix} 
        X_i, Y_i, \Psi_i, V_i
    \end{bmatrix}\trans
    \sim
    \normaldist
    \left(\begin{bmatrix} N_i, E_i, \psi_i, \surgespeed_i \end{bmatrix} \trans, \cov_i\right).
\end{equation}
From \eqref{eq:tcpa} and \eqref{eq:dcpa} it is clear that the output are non-linear transformations of the input parameters. 
The random variable $X_{TCPA}$ corresponding to the \gls{TCPA} between vessels $j$ and $k$, is calculated by replacing the deterministic variables in \eqref{eq:tcpa} with the stochastic variable from \eqref{eq:state_stochastic}. This yields the following definition
\begin{align}\label{eq:tcpa_stochastic}
    X_{TCPA} \triangleq &\alpha \left( (X_j - X_k)(V_j \cos{\Psi_j} - V_k \cos{\Psi_k}) \right. \nonumber \\ 
    &+ \left. (Y_j - Y_k)(V_j \sin{\Psi_j} + V_k \sin{\Psi_k}) \right),
\end{align}
where
\begin{equation}
    \alpha = \frac{1}{V_j^2 + V_k^2 - 2 V_j V_k \cos{\Psi_j - \Psi_k}}.
\end{equation}
Similarly, the expressions for the random varibles $X_{DCPA}$ corresponding to the calculated \gls{DCPA}, and $X_{\bearing}$ corresponding to the calculated bearing $\bearing$ can be defined as
\begin{equation}\label{eq:dcpa_stochastic}
    X_{DCPA} \triangleq \norm{\Delta \mathbf{p}(X_{TCPA})},
\end{equation}
and
\begin{equation}\label{eq:bearing_stochastic}
    X_{\bearing} \triangleq \arctan{\frac{Y_{k} - Y_{j}}{X_{k} - X_{j}}} - \Psi_{j}.
\end{equation}
Deriving a closed-form analytical expression for the density function of $X_{TCPA}$, $X_{DCPA}$ or $X_{\bearing}$ is not a trivial task, if possible at all. 
This is due to the strong non-linearities, the multivariate nature of the transformation and the strong correlation between the random variables in \eqref{eq:state_stochastic}.
Instead, a Monte-Carlo approach is employed to obtain an implicit characterisation of how the uncertainty in the evaluation of the situation will propagate into the distribution of $X_{TCPA}$, $X_{DCPA}$ \& $X_{\bearing}$.
In each scenario an acting agent referred to as \gls{OS} placed at $(N,E) = (0,0)$, with a heading $\course=\ang{0}$ and a forward velocity $\surgespeed = \SI{10}{\meter\per\second}$, with some uncertainty on all four states, such that the initial state of each scenario is drawn from the following distribution
\begin{equation}
    \state_{OS} = \normaldist \left( [0, 0, 0, 10]\trans, \diag{10, 10, 2, 2} \right).
\end{equation}
In each scenario, a target agent referred to as \gls{TV} is placed at a distance of $r=\SI{1000}{\meter}$ from \gls{OS}, for 7 different bearings $\bearing \in \{0, 30, 60, 90, 120, 150, 180\}$.
\begin{equation}
    \state_{TV} \sim \normaldist \left( 
        \begin{bmatrix} 
        r \cos{\frac{\pi}{180}\beta} \\
         r \sin{\frac{\pi}{180}\beta} \\
        180 - \beta \\
        10
    \end{bmatrix}, \diag{10, 10, 2, 2}\right).
\end{equation}
In these examples, the covariance matrices are diagonal matrices, so no cross correlation between the states estimates, however the analysis can similarly be done for correlated estimation noise.
For each bearing/scenario $N=\num{10000}$ samples are drawn and the histograms of the calculated \gls{TCPA}, \gls{DCPA} and $\bearing$ values are visualised in Fig.~\ref{fig:tcpa_hist},  Fig.~\ref{fig:dcpa_hist} and Fig.~\ref{fig:bearing_hist}, respectively.

Looking at the histograms of the \gls{TCPA} values in Fig.~\ref{fig:tcpa_hist}, it would be easy to assume that a gaussian distribution could be fit to the data, when infact performing a Kolmogorov-Smirnov test for any of the distributions will result in a $p$-value of $0$, meaning that the data cannot be described by a gaussian normal distribution.
This makes intuitive sense, as looking back to the definition of the \gls{TCPA} in \eqref{eq:tcpa}, the nonlinear transformation of the gaussian inputs, does not produce a gaussian normal distributed output.

The histograms of the \gls{DCPA} values in Fig.~\ref{fig:dcpa_hist} illustrate the above points even better, as it is clear that depending on the bearing, the family of distribution that could be used is different e.g., for bearing $\bearing=\{30^\circ, 60^\circ, 90^\circ \}$, one could possibly fit a guassian normal distribution, however, for $\bearing = 0^\circ$ a better suited distribution could be either an exponential, Nakagami or Burr distribution.

However, the Kolmogorov-Smirnov (KS) test for any of these distributions will also produce $p$-values of less than 0.05, indicating that these families do not represent the underlying data in a satisfactory manner.
Assuming that a set of known distributions is used to fit the \gls{DCPA} data, where the distribution type used would depend on the region, the fit to the data would naturally be poor in or near the region where the change of distribution would be.
Futhermore, how should these boundaries be chosen? 
Intuitively, one could say that the KS test $p$-value should be used to determine when to switch between distributions, however, the data visualised here is only for 7 different encounters, and will most likely look very different for different random encounters (except for the mirrored versions of the 7 simulated scenarios).
Another way could be to generate random encounters in a grid-search fashion, however, this very quickly becomes intractable as that would be a search space defined in $\R^8$, furthermore, the memory needed to hold all the fitted parameters for each combination of states, would also quickly make this unusable.

\subsection{Kernel Density Estimation}\label{sec:method_kde}

\gls{KDE} is a non-parametric method for estimating the density function purely from data, and is often used when parametric mehthods fail i.e., when no analytical expression of the density function exist, or when fitting of known distributions fimalies yield poor results.

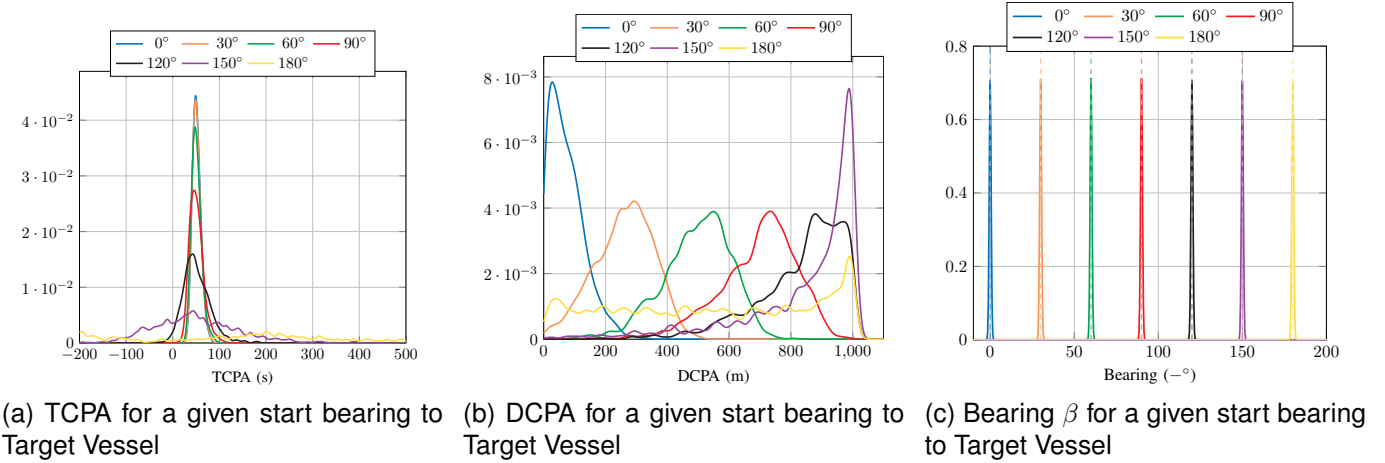
\begin{figure*}[tb]
    \centering
		 \tikzsetnextfilename{tcpa_plot}
    \subfloat[TCPA for a given start bearing to Target Vessel\label{fig:tcpa_plot}]{

        \resizebox{0.31\textwidth}{!}{
            \begin{tikzpicture}
\begin{axis}[
  width=\columnwidth,
  ymin=0,
  xmin=-200,
  xmax=500,
  scaled ticks=false,
  legend style={at={(0.5,1.15)}, anchor=north,legend columns=4},     
  grid,
 xlabel={TCPA (s)}
]
\addplot[no marks,line width=1pt,color=RoyalBlue] table[col sep=comma,x=kernel_tcpa_x, y=kernel_tcpa_y] {data/bearing_0_data.csv};
\addlegendentry{\degrees{0}}

\addplot[no marks,line width=1pt,color=Peach] table[col sep=comma,x=kernel_tcpa_x, y=kernel_tcpa_y] {data/bearing_30_data.csv};
\addlegendentry{\degrees{30}}

\addplot[no marks,line width=1pt,color=Green] table[col sep=comma,x=kernel_tcpa_x, y=kernel_tcpa_y] {data/bearing_60_data.csv};
\addlegendentry{\degrees{60}}

\addplot[no marks,line width=1pt,color=Red] table[col sep=comma,x=kernel_tcpa_x, y=kernel_tcpa_y] {data/bearing_90_data.csv};
\addlegendentry{\degrees{90}}

\addplot[no marks,line width=1pt,color=Black] table[col sep=comma,x=kernel_tcpa_x, y=kernel_tcpa_y] {data/bearing_120_data.csv};
\addlegendentry{\degrees{120}}

\addplot[no marks,line width=1pt,color=Purple] table[col sep=comma,x=kernel_tcpa_x, y=kernel_tcpa_y] {data/bearing_150_data.csv};
\addlegendentry{\degrees{150}}

\addplot[no marks,line width=1pt,color=Goldenrod] table[col sep=comma,x=kernel_tcpa_x, y=kernel_tcpa_y] {data/bearing_180_data.csv};
\addlegendentry{\degrees{180}}

\end{axis}
\end{tikzpicture}
        }
    }
    \hfill
	\tikzsetnextfilename{dcpa_plot}
    \subfloat[DCPA for a given start bearing to Target Vessel\label{fig:dcpa_plot}]{
        \resizebox{0.31\textwidth}{!}{
            \begin{tikzpicture}
\begin{axis}[
  width=\columnwidth,
  ymin=0,
  xmin=0,
  xmax=1100,
  scaled ticks=false,
  legend style={at={(0.5,1.15)}, anchor=north,legend columns=4},     
  grid,
 xlabel={DCPA (m)}
]
\addplot[no marks,line width=1pt,color=RoyalBlue] table[col sep=comma,x=kernel_dcpa_x, y=kernel_dcpa_y] {data/bearing_0_data.csv};
\addlegendentry{\degrees{0}}

\addplot[no marks,line width=1pt,color=Peach] table[col sep=comma,x=kernel_dcpa_x, y=kernel_dcpa_y] {data/bearing_30_data.csv};
\addlegendentry{\degrees{30}}

\addplot[no marks,line width=1pt,color=Green] table[col sep=comma,x=kernel_dcpa_x, y=kernel_dcpa_y] {data/bearing_60_data.csv};
\addlegendentry{\degrees{60}}

\addplot[no marks,line width=1pt,color=Red] table[col sep=comma,x=kernel_dcpa_x, y=kernel_dcpa_y] {data/bearing_90_data.csv};
\addlegendentry{\degrees{90}}

\addplot[no marks,line width=1pt,color=Black] table[col sep=comma,x=kernel_dcpa_x, y=kernel_dcpa_y] {data/bearing_120_data.csv};
\addlegendentry{\degrees{120}}

\addplot[no marks,line width=1pt,color=Purple] table[col sep=comma,x=kernel_dcpa_x, y=kernel_dcpa_y] {data/bearing_150_data.csv};
\addlegendentry{\degrees{150}}

\addplot[no marks,line width=1pt,color=Goldenrod] table[col sep=comma,x=kernel_dcpa_x, y=kernel_dcpa_y] {data/bearing_180_data.csv};
\addlegendentry{\degrees{180}}

\end{axis}
\end{tikzpicture}
        }
    }
    \hfill
    \tikzsetnextfilename{bearing_plot}
    \subfloat[Bearing $\bearing$ for a given start bearing to Target Vessel\label{fig:bearing_plot}]{
        \resizebox{0.31\textwidth}{!}{
            \begin{tikzpicture}
\begin{axis}[
  width=\columnwidth,
  ymin=0,
  xmin=-10,
  xmax=200,
  ymax=0.8,
  legend style={at={(0.5,1.15)}, anchor=north,legend columns=4},     
  grid,
  xlabel={Bearing (\degrees{-})}
]

\addplot[no marks,line width=1pt,color=RoyalBlue] table[col sep=comma,x=kernel_bearing_x, y=kernel_bearing_y] {data/bearing_0_data.csv};
\addplot[no marks,line width=1pt,color=Peach] table[col sep=comma,x=kernel_bearing_x, y=kernel_bearing_y] {data/bearing_30_data.csv};
\addplot[no marks,line width=1pt,color=Green] table[col sep=comma,x=kernel_bearing_x, y=kernel_bearing_y] {data/bearing_60_data.csv};
\addplot[no marks,line width=1pt,color=Red] table[col sep=comma,x=kernel_bearing_x, y=kernel_bearing_y] {data/bearing_90_data.csv};
\addplot[no marks,line width=1pt,color=Black] table[col sep=comma,x=kernel_bearing_x, y=kernel_bearing_y] {data/bearing_120_data.csv};
\addplot[no marks,line width=1pt,color=Purple] table[col sep=comma,x=kernel_bearing_x, y=kernel_bearing_y] {data/bearing_150_data.csv};
\addplot[no marks,line width=1pt,color=Goldenrod] table[col sep=comma,x=kernel_bearing_x, y=kernel_bearing_y] {data/bearing_180_data.csv};

\addlegendentry{\degrees{0}}
\addlegendentry{\degrees{30}}
\addlegendentry{\degrees{60}}
\addlegendentry{\degrees{90}}
\addlegendentry{\degrees{120}}
\addlegendentry{\degrees{150}}
\addlegendentry{\degrees{180}}

\addplot[thick, dashed, samples=50, smooth,domain=0:6,RoyalBlue!50] coordinates {(0,0)(0,1)};
\addplot[thick, dashed, samples=50, smooth,domain=0:6,Peach!50] coordinates {(30,0)(30,1)};
\addplot[thick, dashed, samples=50, smooth,domain=0:6,Green!50] coordinates {(60,0)(60,1)};
\addplot[thick, dashed, samples=50, smooth,domain=0:6,Red!50] coordinates {(90,0)(90,1)};
\addplot[thick, dashed, samples=50, smooth,domain=0:6,Black!50] coordinates {(120,0)(120,1)};
\addplot[thick, dashed, samples=50, smooth,domain=0:6,Purple!50] coordinates {(150,0)(150,1)};
\addplot[thick, dashed, samples=50, smooth,domain=0:6,Goldenrod!50] coordinates {(180,0)(180,1)};

\end{axis}
\end{tikzpicture}
        }
    }
    \caption{\gls{KDE} of \gls{TCPA}, \gls{DCPA} and bearing values for the 7 different scenarios.}
    \label{fig:kde_estimates}
\end{figure*}

Let $\hat{f}_h(x)$ be the estimated density function, given by
\begin{equation}\label{eq:kde_function}
    \hat{f}_h(x) = \frac{1}{nh} \sum_{i=0}^{n} K \left(\frac{x - x_i}{h} \right),
\end{equation}
where $n$ is the number of data points, $K$ is the kernel function, and $h$ is a smoothing parameter called the \emph{bandwidth}.
A widely used kernel function is the Gaussian Kernel function.
\begin{equation}\label{eq:gaussian_kernel}
    K(x, x_i) = \exp \left\{ \frac{\norm{x - x_i}^2}{2h^2} \right\},
\end{equation}
inserting \eqref{eq:gaussian_kernel} into \eqref{eq:kde_function} yields
\begin{equation}\label{eq:kde_gaussian}
    \hat{f}_h(x) = \frac{1}{nh} \sum_{i=1}^{n} \exp \left\{ \frac{\norm{x - x_i}^2}{2h^2} \right\}.
\end{equation}
Applying \eqref{eq:kde_gaussian} to the \gls{TCPA}/\gls{DCPA} data with bandwidth $h_{TCPA}=5$, $h_{DCPA}=15$ and $h_{\bearing} = 0.5$ (chosen arbitrarily for illustrative purposes) results in the density functions shown in Fig.~\ref{fig:tcpa_plot} and Fig.~\ref{fig:dcpa_plot} respectively.
Fig.~\ref{fig:tcpa_plot} and Fig.~\ref{fig:dcpa_plot} show the \gls{KDE} of \gls{TCPA} and \gls{DCPA} respectively.
Fig.~\ref{fig:dcpa_plot} clearly illustrates how the different scenarios would need require vastly different distributions, if parametric methods were employed to fit the \gls{DCPA} values.

\begin{figure}[tb]
	\centering
	\tikzsetnextfilename{kde_bandwidth_estimation}
		\resizebox{\columnwidth}{!}{\begin{tikzpicture}
    \begin{axis} [
        width=\columnwidth,
        scaled ticks=false,
        legend style={at={(0.5,1.15)}, anchor=north,legend columns=4},     
        grid,
        xmin=-50,
        xmax=650,
        ylabel={Density},
        xlabel={DCPA (m)}
    ]
    \addplot[ybar,fill,width=1.5pt,color=RoyalBlue!40, hist={density, bins=300}] table [y=dcpa, col sep=comma] {data/kde_dcpa.csv};
    \addlegendentry{Data}

    \addplot[no marks,line width=1.2pt,color=Peach] table[col sep=comma,x=x, y=ISJ] {data/kde_bandwidth_estimation.csv};
    \addlegendentry{$\hat{f}_{ISJ}$}

    \addplot[no marks,line width=1.2pt,color=RoyalBlue] table[col sep=comma,x=x, y=silverman] {data/kde_bandwidth_estimation.csv};
    \addlegendentry{$\hat{f}_{silverman}$}

    \addplot[no marks,line width=1.2pt,color=Red] table[col sep=comma,x=x, y=grid_search] {data/kde_bandwidth_estimation.csv};
    \addlegendentry{$\hat{f}_{grid}$}

    \end{axis}
\end{tikzpicture}}
	\caption{\gls{KDE} estimates of $n=10^5$ samples of \gls{DCPA} using three different methods for estimating the bandwidth: \gls{ISJ}, Silverman's rule of thumb and lastly a grid-search based method using cross-validation.}
	\label{fig:kde_bandwidth_estimation}
\end{figure}
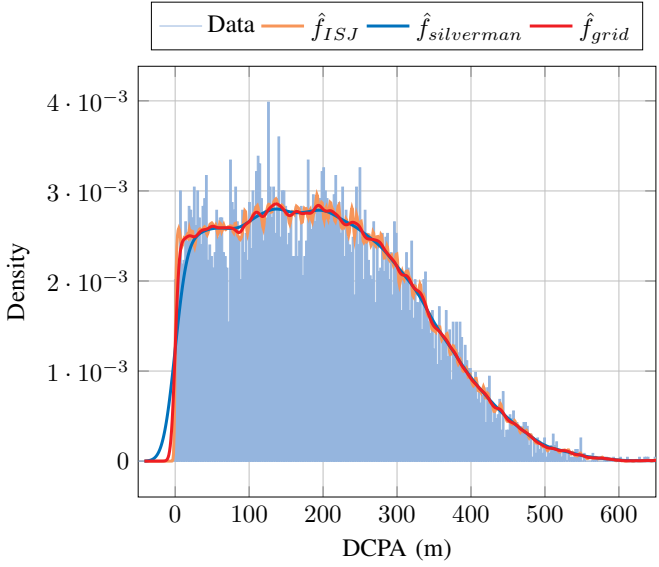

When using \gls{KDE} based methods the choice of bandwidth $h$ is important and often more so than the choice of kernel, with several methods availble for choosing the best bandwidth including; grid-search with cross-validation and \emph{Silverman's rule of thumb} \cite{Silverman1996}.
Silverman's rule is a fast algorithm for estimating the correct bandwidth, however the rule assumes that the true underlying density is a Gaussian normal distribution, which does not apply for the data transformed through \eqref{eq:tcpa_stochastic}, \eqref{eq:dcpa_stochastic} or \eqref{eq:bearing_stochastic}.
When dealing with non-normal data the \gls{ISJ} algorithm \cite{Botev2010} can be used, as this algorithm does not assume an underlying normal distribution in the data, the trade-off is that it requires more data for a good estimate, thereby increasing the computational time.
From the \gls{DCPA} data of a randomly generated scenario (shown in , three different bandwidth are estimated; $h_{grid}$, $h_{silverman}$ and $h_{ISJ}$, based on grid-search cross validated, Silvermans rule of thumbs and the \gls{ISJ} algorithm respectively.
The grid-search is over the range $h_{grid} \in [0.05, 12]$ with increments of $\Delta = 0.05$.
The resulting bandwidth estimates are
\begin{align*}
    h_{grid} &= 4.5 \\
    h_{silverman } &= 12.812 \\
    h_{ISJ} &= 1.169,
\end{align*}
Fig.~\ref{fig:kde_bandwidth_estimation} visualises how the different approaches compare with the true underlying density.
From the figure it is evident that the $h_{silverman}$ is too wide of a bandwidth, as it fails to correctly capture the sharp transition around $\mathrm{DCPA}=0m$, this is also expected as the underlying data is not generated from a normal distribution.
The smaller values of $h_{ISJ}$ and $h_{grid}$ result in quite similar density estimates, with $h_{ISJ}$ having a slightly better estimate around $\mathrm{DCPA}=0m$, albeit with a more variance in the rest of the region.
So which method should be used? 
Clearly bandwidth estimates based on silverman's rule of thumb should not be used, as these fail to capture the underlying non-gaussianity, leaving the grid-search and \gls{ISJ} based methods as viable methods.
The grid-search based method is inherently slow, and the example shown takes around 5 minutes running on a 48 core AMD Threadripper 3960X CPU, compared to the very fast FFT-based implementation of the \gls{ISJ} in \cite{Odland2018} which takes less than a second on the same hardware.
One could argue that once the bandwidth is estimated through a grid-search, the value $h_{grid}$ could be used for all future evaluations, however, the high non-linearities in \eqref{eq:tcpa_stochastic}, \eqref{eq:dcpa_stochastic} and \eqref{eq:bearing_stochastic} does not guarantee that this choice would be valid in any scenario.
Comparing to the computational cost of drawing more samples for the \gls{ISJ} algorithm and re-fitting the \gls{KDE} every time, to the computational cost of having to run a grid-search, the \gls{ISJ} algorithm is the favorable approach.
The resulting \gls{KDE} based algorithm for the stochastic behavioural model is outlined in Algorithm \ref{alg:kde_algorithm}.
\begin{algorithm}[tb]
    \caption{Algorithm for estimating probability of collision risk, and probability of \gls{OS} give-way obligation}\label{alg:kde_algorithm}
    \begin{algorithmic}[1]
	        \Require Estimated states $\eststate_{1}, \eststate_{2}$, Covariances $\cov_1, \cov_2$, number of samples $N$, comfort zone radius $d_{act}$.
	        \Ensure $\probability{\text{``risk"}}$, $\probability{\text{``give-way"}}$.
	        \State $B_{TCPA}[1:N], B_{DCPA}[1:N], B_{\bearing1}[1:N], B_{\bearing2}[1:N], B_{\Delta \course}[1:N] \gets 0$ \Comment{Create empty buffers of size $N$.}
	        \State $P_1, P_2 \gets 0$  \Comment{Empty situation evaluation arrays for each vessel.}
	        \For {$i = 1$ to $N$}.
         		\State $\state_{1,i} \sim \normaldist(\eststate_1, \cov_1)$  \Comment{Draw sample states.}
	            \State $\state_{2,i} \sim \normaldist(\eststate_2, \cov_2)$
	            \State $\regtext{TCPA}_i, \regtext{DCPA}_i \gets \regtext{CalcTcpaDcpa}(\state_{1,i}, \state_{2,i})$ \Comment{Calc. TCPA/DCPA for sample $i$.}
	            \State $\bearing_{1, i}^2, \bearing_{2, i}^1 \gets \regtext{CalcRelBearings}(\state_{1,i}, \state_{2,i})$ 
	            \State $\Delta \course_i \gets \regtext{CalcReciprocalCourse}(\state_{1,i}, \state_{2,i})$
 	            \State $B_{TCPA}[i], B_{DCPA}[i] \gets \regtext{TCPA}_i, \regtext{DCPA}_i$  \Comment{Store in buffers.}
 	            \State $B_{\bearing1}[i], B_{\bearing2}[i] \gets \bearing_{1, i}^2, \bearing_{2, i}^1$
 	            \State $B_{\Delta \course}[i] \gets \Delta \course_i$ 	            

	        \EndFor
	        \State $\hat{f}_{TCPA}$, $\hat{f}_{DCPA}$, $\hat{f}_{\bearing1}$, $\hat{f}_{\bearing2}, \hat{f}_{\Delta \course} \gets \regtext{FitPDFs}(B_{TCPA}, B_{DCPA}, B_{\bearing1}, B_{\bearing2}, B_{\Delta \course})$
	        \State $\probability{\text{``risk"}} \gets \regtext{CalcRisk}(\hat{f}_{DCPA})$ \Comment{Eq. \eqref{eq:prob_dcpa}.}
	
	        \For {$i=1$ to $2$}  \Comment{Map the situations for each vessel.}
	            \State $v \gets g_1(\bearing_i)$ 
	            \State $P_{HO, i} \gets \probability{v = HO}$ 
	            \State $P_{SB, i} \gets \probability{v = SB}$ 
	            \State $P_{OT, i} \gets \probability{v = OT}$ 
	            \State $P_{PS, i} \gets \probability{v = PS}$ 
	            \State $P_i\gets \{P_{HO, i}, P_{SB, i}, P_{OT, i}, P_{PS, i}\}$ \Comment{Store probabilities for vessel $i$.}
	        \EndFor
	        \State $G \gets g_2(P_1, P_2)$ \Comment{Calc. give-way set from Eq. \eqref{eq:give_way_set}.}
	        \State $\probability{\text{``give-way"}} \gets \probability{\text{``risk"}} \cdot \sum\limits_{g \in G} g$ \Comment{Calc. total probability of give-way.} \\
	        \Return $\probability{\text{``risk"}}$, $\probability{\text{``give-way"}}$ 
	    \end{algorithmic} 
\end{algorithm}
A \gls{KDE} based approach has provide a simple yet effective way of estimating highly non-linear transformations of \gls{TCPA}, \gls{DCPA} and bearing, but what are the drawbacks of this method?
Two things come to mind; first the importance of choosing the right bandwidth is crucial for a correct estimate of the density function, and while the \gls{ISJ} algorithm has shown satisfactory results for the cases illustrated, due to the high non-linearities of the \gls{TCPA}/\gls{DCPA} equations, it is impossible to say that it might be a good fit for all possible scenarios.
The second thing to note is that the \gls{KDE} requires storing all the original data in memory, as this is used run-time when evaluating the density of at a given point.
This might not be a problem for a single target, however in an actual setting where the number of targets can easily surpass 30-50 targets, this might be an issue in a onboard system running many different applications, where each application is required to have as small a memory footprint as possible.
Lastly, while not a flaw of the \gls{KDE} method itself, it should be noted that the use of \gls{KDE} based methods, limits the use of arbitrarily shaped comfort-zones, as the distance based \gls{DCPA} requires that there exists a function to transform the chosen shape into a circle, for the $L_2$ norm $||\cdot||$ to be applicable.

\subsection{Discrete Event Systems}

Another method for interpreting a situation in a probabilistic manner is the use of \gls{DES}, so-called automata, more specifically stochastic automata \cite{Cassandras2008}.
A stochastic automaton $\mathcal{G}_s$ is defined by the five-tuple \cite{Blanke2015}
\begin{equation}\label{eq:stochastic_automata_definition}
    \mathcal{G}_s = (\mathcal{Z}, \mathcal{V}, \mathcal{W}, L, p_0),
\end{equation}
where $\mathcal{Z}$ is the set of possible states, $\mathcal{V}$ is the set of possible inputs, $\mathcal{W}$ is the set of possible outputs of the automaton, $p_0$ is the initial state probability distribution.
Finally, $L$ is the \emph{behavioural relation} of the automaton, and is a mapping from $L: \mathcal{Z} \times \mathcal{W} \times \mathcal{Z} \times \mathcal{V} \rightarrow [0, 1]$.
The probability distributions also have the following properties
\begin{align}\label{eq:behavioral_relation}
    0 \leq  &L(z^{\prime}, w |z, v) \leq 1, \quad \forall z \in \mathcal{Z}, v \in \mathcal{V}, w \in \mathcal{W}, \\
    \sum_{z'\in \mathcal{Z}} \sum_{w \in \mathcal{W}} &L(z', w |z, v) = 1, \quad \forall z \in \mathcal{Z}, v \in \mathcal{V},
\end{align}
which leads to
\begin{align}\label{eq:automata_transition_probability}
    G(z'|z, v) &= \sum_{w \in \mathcal{W}} L(z', w | z, v), \\
    H(w| z, v) &= \sum_{z' \in \mathcal{Z}} L(z', w | z, v),
\end{align}
where $G(z'| z, v)$ is called the \emph{state transition relation} i.e. the state transition probability, and $H(w|z, v)$ is the \emph{output relation} of the stochastic automata \cite{Blanke2015}.
\begin{figure}[tb]
	\centering
    \tikzsetnextfilename{understanding_automata}
    \resizebox{\columnwidth}{!}{    
        \includegraphics{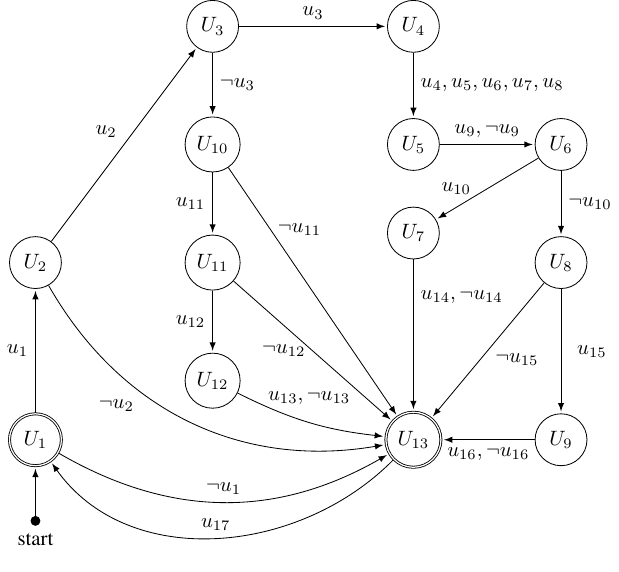}
    }
	\caption{State transition diagram of the understanding automaton defined in \cite{Hansen2020}.}\label{fig:understanding_automata}
\end{figure}
The \emph{understanding automaton} \cite{Papageorgiou2022} illustrated in Fig.~\ref{fig:understanding_automata} implements the understanding part of the situational awareness \cite{Endsley1995}, and is able to analyse a given situation between two vessels, by a interpreting the raw state estimates of two vessels, producing a higher level abstraction which describes the situation in human-interpretable language, thus the correct \gls{COLREGs} rule can be applied accordingly.
For further details of the understanding automaton, including all state definitions and output words, the reader is referred to the work presented in \cite{Papageorgiou2019, Hansen2020, Papageorgiou2022}.
However, as each state in understanding automaton performs a specific test on some already existing data, the automaton does not have inputs in the classical sense i.e., there are no "underlying process" which is being modelled.
This means that a stochastic implementation of the understanding automaton differs from the 5-tuple definition given in \eqref{eq:stochastic_automata_definition}, as the set of inputs $\mathcal{V}$ belongs to the empty set
\begin{equation}
    \mathcal{V} = \emptyset,
\end{equation}
as such, it follows that $L$ is defined as the mapping
\begin{equation}
    L: \mathcal{Z} \times \mathcal{W} \times \mathcal{Z} \rightarrow [0, 1],
\end{equation}
with 
\begin{align}\label{eq:behavioral_relation_new}
    L(z^{\prime}, w, z) = \condprob{z^{\prime}, w}{z}, \quad z, z^{\prime} \in \mathcal{Z}, w \in \mathcal{W}.
\end{align}
The focus in this study is on the effect of the tracking uncertainty of a given target, namely $\eststate$, under the assumption that the type of a tracked target (buoy, power-driven vessel, sailboat, etc.) is known.
According to \eqref{eq:behavioral_relation_new}, the tracking uncertainty affects the behavioral relation corresponding to the states of the set $\mathcal{U}^\prime \subset \mathcal{U}$ defined as
\begin{equation}
    \mathcal{U}^\prime \triangleq \set{U_1, U_2, U_4, U_8, U_9},
\end{equation}
where $\mathcal{U}$ is the set of all states of the understanding automaton \cite{Papageorgiou2022}.
\begin{remark}\normalfont
    According to the definition of the understanding automaton, the output of the states $U_1, U_8$ are determined by the violation of threshold values ($d_{aware}, d_{act}$), in relation to the \gls{DCPA} calculation \cite{Papageorgiou2022}.
    Similarly, the output of $U_2, U_9$ are determined by the violation of \gls{TCPA} threshold values ($t_{aware}, t_{act}$).
    Lastly, the output of $U_4$ depends on the relative bearing mapping given in \eqref{eq:bearing_colregs_map}.
\end{remark}

The transition probabilities for the states in $\mathcal{U}^\prime$ can be determined by the density functions $f_{TCPA}$, $f_{DCPA}$ and $f_{\bearing}$.
These density functions can be estimated by use of \gls{KDE} methods, as described earlier.
However, this would have the same memory requirements, which might not be desirable.
The solution is to ``sample" from the distributions of the input parameters in the states of the understanding automaton, transform the samples through the \gls{TCPA}, \gls{DCPA} and bearing equations, and based on the resulting values, trigger the appropriate transitions.
The understanding automaton is designed in such a way that transitions to the marked states $U_1$ and $U_{13}$ will always happen, irrespective of the inputs, i.e. all inputs will cause the automaton to perform a full loop, ending with the transition to the marked and final state $U_1$ from state $U_{13}$.
Each time the automaton has transitioned through all the states and ends up in the marked state $U_1$, the output caused by the transitions can be concatenated into a string $s$, 
This ensures that the automaton output produces a \emph{string} that can be mapped to a unique situation e.g., "No risk", "Port Crossing", "Overtaking" or "Head-on".

As an example, the state $U_{8}$ will output $u_{15}$ if $DCPA \leq d_{act}$, and $\lnot u_{15}$ otherwise, from this follows that the  probability of a collision risk, given $N$ strings $[s_1, s_2, \ldots, s_N]$ from the understanding automaton is
\begin{equation}
    \probability{DCPA \leq d_{act}} = \frac{1}{N} \sum\limits_{i=1}^N H_{d_{act}}(s_i),
\end{equation}
where 
\begin{equation}
    H_{d_{act}}(s) = \begin{cases}
        1, \quad \text{if } u_{15} \in s,\\
        0, \quad \text{otherwise}.
    \end{cases}
\end{equation}
Similar probabilities can be calculated for the situation types such that %
\begin{align}
    \probability{E_i} &= \frac{1}{N} \sum\limits_{i=1}^N H_{E_i}(s_i),
\end{align}
where  $E_i \in \{(R_0, 1), (R_{13}, 0), (R_{13}, 1), (R_{14}, 1), (R_{15}, 0), (R_{15}, 1)\}$ (elements in $\mathcal{S}_2$) and
\begin{align}
    H_{(R_0, 1)}(s) &= \begin{cases}
        1, \quad \text{if } \{u_{4}, u_{5}, u_{6}, u_{7}, u_{8}\} \not\in s,\\
        0, \quad \text{otherwise},
    \end{cases} \\
    H_{(R_{13}, 0)}(s) &= \begin{cases}
        1, \quad \text{if } u_{4} \in s,\\
        0, \quad \text{otherwise},
    \end{cases} \\
    H_{(R_{13}, 1)}(s) &= \begin{cases}
        1, \quad \text{if } u_{5} \in s,\\
        0, \quad \text{otherwise},
    \end{cases} \\
    H_{(R_{14}, 1)}(s) &= \begin{cases}
        1, \quad \text{if } u_{6} \in s,\\
        0, \quad \text{otherwise},
    \end{cases} \\
    H_{(R_{15}, 0)}(s) &= \begin{cases}
        1, \quad \text{if } u_{7} \in s,\\
        0, \quad \text{otherwise},
    \end{cases} \\
    H_{(R_{15}, 1)}(s) &= \begin{cases}
        1, \quad \text{if } u_{8} \in s,\\
        0, \quad \text{otherwise}.
    \end{cases}
\end{align}
It should be noted that the evaluation of $H_{(R_0, 1)}(s)$ (no \gls{COLREGs} rule applies) requires that the state $U_4$ does not produce an output, i.e. it produces an empty string during the transition to $U_5$.
It then follows that the total probability that \gls{OS} needs to give way is
\begin{align}
    \probability{\text{``give-way"}} &= \probability{(R_0, 1)} + \probability{(R_{13}, 1)}  \nonumber \\ 
    &+ \probability{(R_{14}, 1)}+ \probability{(R_{15}, 1)}
\end{align}
The resulting automata based algorithm for the stochastic behavioural model is outlined in Algorithm \ref{alg:des_algorithm}.
\begin{algorithm}[t]
	\caption{Algorithm for estimating probability of collision risk, and probability of \gls{OS} give-way obligation using automata}\label{alg:des_algorithm}
	\begin{algorithmic}[1]
		\Require Estimated states $\eststate_{1}, \eststate_{2}$, Covariances $\cov_1, \cov_2$, number of samples $N$, comfort zone radius $d_{act}$	
		\Ensure Collision probability $\probability{C_r}$, Give-way probability $\probability{\text{``give-way"}}$
		\State $c_{act}, c_{R_0, 1}, c_{R_{13}, 0}, c_{R_{13}, 1}, c_{R_{14}, 1}, c_{R_{15}, 0}, c_{R_{15}, 1} \gets 0$ \Comment{Initialise counters.}
		
		\For {$i = 1$ to $N$}
		\State $\state_{1,i} \sim \normaldist(\eststate_1, \cov_1)$  \Comment{Draw sample states.}
		\State $\state_{2,i} \sim \normaldist(\eststate_2, \cov_2)$
		\State $s \gets G_u(\state_{1,i}, \state_{2,i}, d_{act})$  \Comment{Generate the string $s$ from drawn samples}
		
		\If{$u_{15} \in s$} \Comment{Samples result in collision risk.}
		\State $c_{act} \gets c_{act}  + 1$ \Comment{Increment counter.}
		\EndIf
		
		\If{$u_4 \in s$}  \Comment{Check situation type and increment relevant counter.}
		\State $c_{R_{13}, 0} \gets c_{R_{13}, 0} + 1$
		\ElsIf{$u_5 \in s$}
		\State $c_{R_{13}, 1} \gets c_{R_{13}, 1} + 1$
		\ElsIf{$u_6 \in s$}
		\State $c_{R_{14}, 1} \gets c_{R_{14}, 1} + 1$
		\ElsIf{$u_7 \in s$}
		\State $c_{R_{15}, 0} \gets c_{R_{15}, 0} + 1$
		\ElsIf{$u_8 \in s$}
		\State $c_{R_{15}, 1} \gets c_{R_{15}, 1} + 1$
		\Else
		\State $c_{R_0, 1} \gets c_{R_0, 1} + 1$  \Comment{No situation, map to $R_0$.}
		\EndIf
		\EndFor
		
		\State $\probability{C_r} \gets \frac{1}{N} c_{act}$  \Comment{Calculate collision prob. }
		\State $\probability{\text{``give-way"}} \gets \frac{1}{N} (c_{R_0, 1} + c_{R_{13}, 1} + c_{R_{14}, 1} + c_{R_{15}, 1} + c_{R_0, 1}) \cdot \probability{C_r}$ \\
		\Return $\probability{\text{``risk"}}$, $\probability{\text{``give-way"}}$ 
	\end{algorithmic} 
\end{algorithm}

\section{Experiments}\label{sec:experiements}

\begin{figure*}[!t]
    \centering
    \subfloat[Scenario 1: Starboard crossing\label{fig:sim_starboard}]{
        \resizebox{0.31\textwidth}{!}{
            \tikzsetnextfilename{sim_starboard}
                \includegraphics{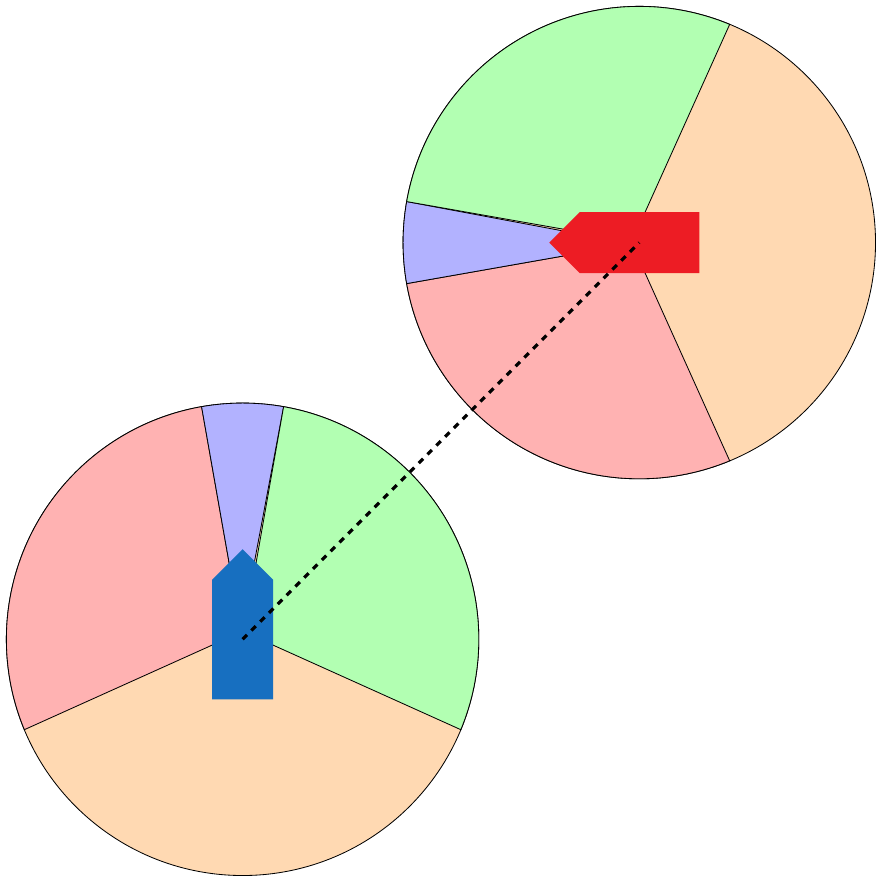}
        }
    }
    \hfill
    \subfloat[Scenario 2: Port/Head-on\label{fig:sim_port_headon}]{
        \resizebox{0.21\textwidth}{!}{
            \tikzsetnextfilename{sim_port_headon}
                \includegraphics{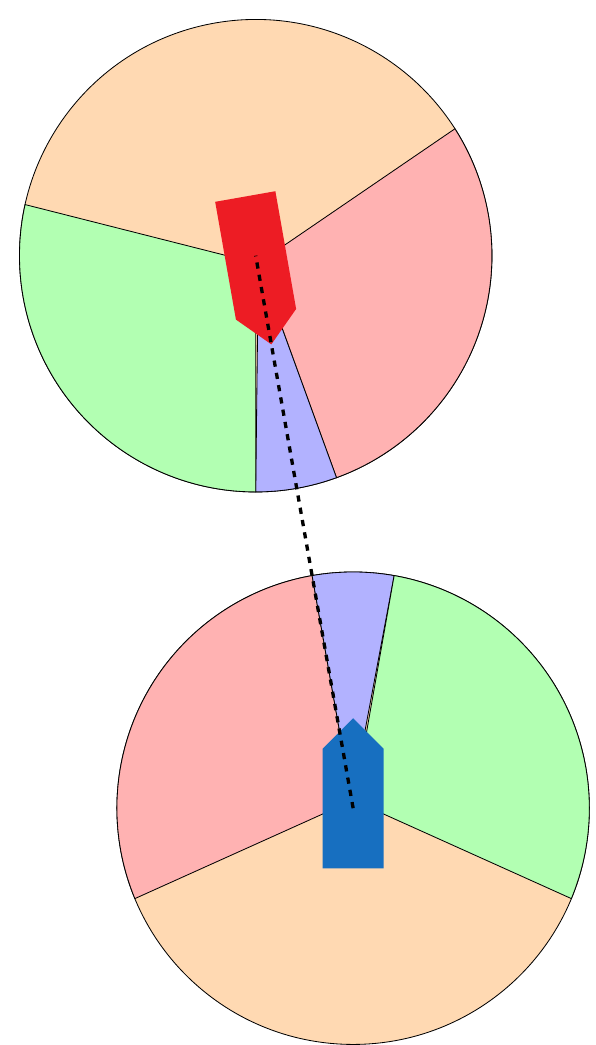}
        }
    }
    \hfill
    \tikzsetnextfilename{bearing_plot}
    \subfloat[Scenario 3: Overtaking/Port crossing\label{fig:sim_overtaking_starboard}]{
        \resizebox{0.31\textwidth}{!}{
        \tikzsetnextfilename{sim_overtaking_starboard}
                \includegraphics{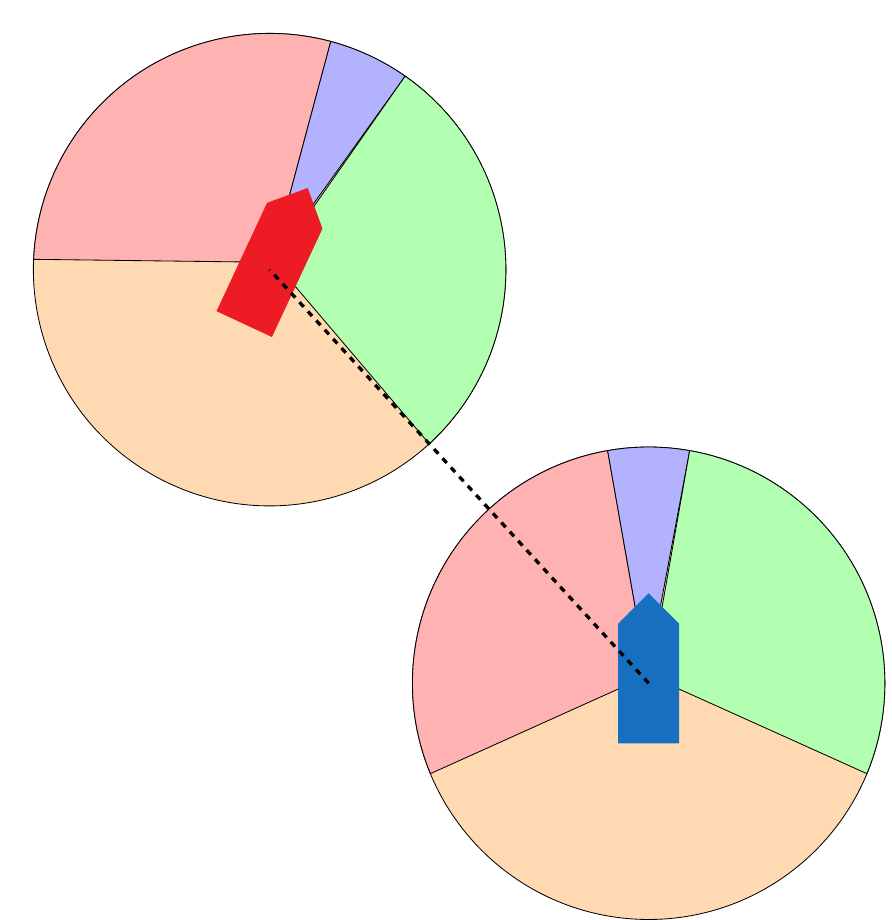}
        }
    }
    \caption{Illustrations of the three different simulation scenarios. \gls{OS} is denoted as a grey circle with a black circle inside, and \gls{TV} is denoted as a black dot.}
    \label{fig:sim_scenarios}
\end{figure*}

To illustrate how the uncertainty can effect the decision-making in an autonomous system, and how the \gls{KDE} based method compares against the automaton based method, three different scenarios are simulated;
\begin{enumerate}
    \item Starboard crossing
    \item Head-on/Port crossing 
    \item Overtaking/Port crossing
\end{enumerate}
The starboard crossing scenario will focus on the risk of collision, whereas the remaining two scenarios will analyse the two boundaries where the \gls{COLREGs} obligation for \gls{OS} will switch.
The different scenarios are illustrated in Fig.~\ref{fig:sim_scenarios}.
All scenarios have a \gls{OS} of ship length $l_{OS}=\SI{30}{\meter}$, with a comfort zone 5 times larger than the ship length, such that
\begin{equation}
    d_{act} = 5 \cdot l_{OS} = \SI{150}{\meter},
\end{equation}
and \gls{OS} is located at $(N,E) = (0, 0)$ with a forward velocity $\surgespeed = 10 \frac{m}{s}$.
For simplicity the simulations assume perfect knowledge of the state $\state_{OS}$, meaning the covariance matrix $\Sigma_{OS}$ is
\begin{equation}
    \Sigma_{OS} = \mathbf{0},
\end{equation}
however, the method can accommodate non-zero covariance matrices of \gls{OS} just as easily.
Each scenario contains a single \gls{TV}, where the tracking error $\Sigma_{TV}$ will be evaluated a six different levels of uncertainty such that
\begin{equation}
    \Sigma_{TV} = \alpha \cdot \diag{10, 10, 2, 2},
\end{equation}
where $\alpha \in \{0.1, 0.5, 1.0, 1.5, 2.0, 5.0\}$, whereas the true state $\state_{TV}$ will differ for each scenario.
The maximum value of $\alpha=5.0$ results in a high uncertainty (more so for the speed and course estimates than the position estimates), however these extreme values should provide better insight into how the method performs overall.
For each simulation $N=100,000$ samples will be drawn for the \gls{KDE}, and the automaton will be evaluated for the same number of iterations.

\subsection{Starboard Crossing}

This scenario has \gls{TV} placed at $(N,E) = (1250, 1000)$, with a course $\course = \ang{270}$ and a speed $\surgespeed = \SI{10}{\meter\per\second}$.
The scenario is illustrated in Fig.~\ref{fig:sim_starboard}.
In this scenario the obligation of \gls{OS} in case of a collision is clear, \gls{COLREGs} Rule 15 applies, and \gls{OS} should give way.
However, the situation without uncertainty yields a $DCPA = \SI{176.78}{\meter} \nleq \SI{150}{\meter}$, so the situation does not pose a risk.
Table \ref{tab:sim_scenario_1} shows the resulting simulations of the \gls{KDE} based algorithm (denoted ``KDE"), and the automata based algorithm (denoted ``DES") at varying levels of uncertainty $\alpha$, where the notation $\probability{d_{act}}$, $\probability{R_{X}}$, $\probability{\text{``give-way"}}$ should be interpreted as ``probability of collision", ``probability of Rule X" and ``probability that \gls{OS} needs to give way", respectively.
\begin{table*}[tb]
    \centering
    \caption{Analysis of scenario 1: starboard crossing}
    \label{tab:sim_scenario_1}
    \begin{tabular}{r|c c c c c c c c c c c c}
        \toprule
         & \multicolumn{2}{c}{$\probability{d_{act}}$} & \multicolumn{2}{c}{$\probability{R_{0}}$} & \multicolumn{2}{c}{$\probability{R_{13}}$} & \multicolumn{2}{c}{$\probability{R_{14}}$} & \multicolumn{2}{c}{$\probability{R_{15}}$} & \multicolumn{2}{c}{$\probability{\text{``give-way"}}$} \\
        \cmidrule(lr){2-3}
        \cmidrule(lr){4-5}
        \cmidrule(lr){6-7}
        \cmidrule(lr){8-9}
        \cmidrule(lr){10-11}
        \cmidrule(lr){12-13}
        $\alpha$ & KDE & DES & KDE & DES & KDE & DES & KDE & DES & KDE & DES & KDE & DES \\
        \midrule
        0.1 & 0.052 & 0.051 & 0.000 & 0.000 & 0.000 & 0.000 & 0.000 & 0.000 & 1.000 & 1.000 & 0.052 & 0.051 \\
        0.5 & 0.371 & 0.371 & 0.000 & 0.000 & 0.000 & 0.000 & 0.000 & 0.000 & 1.000 & 1.000 & 0.371 & 0.371 \\
        1.0 & 0.395 & 0.394 & 0.000 & 0.000 & 0.000 & 0.000 & 0.000 & 0.000 & 1.000 & 1.000 & 0.395 & 0.394 \\
        1.5 & 0.331 & 0.333 & 0.000 & 0.000 & 0.000 & 0.000 & 0.000 & 0.000 & 1.000 & 1.000 & 0.331 & 0.333 \\
        2.0 & 0.274 & 0.275 & 0.000 & 0.000 & 0.000 & 0.000 & 0.000 & 0.000 & 1.000 & 1.000 & 0.273 & 0.275 \\
        5.0 & 0.129 & 0.130 & 0.000 & 0.000 & 0.000 & 0.000 & 0.000 & 0.000 & 1.000 & 1.000 & 0.129 & 0.130 \\
        \bottomrule
    \end{tabular}
\end{table*}
From Table \ref{tab:sim_scenario_1}, it is clear the autonomous system sees a clear Rule 15 situation, as expected.
However, the $\probability{d_{act}}$ increases to a maximum of $\probability{d_{act}}=0.395$ at $\alpha = 1.0$, and then decreases as $alpha$ continues to increase, which directly translates to the same behaviour in the ``give-way" probability.
This is to be expected, as with increasing uncertainty, the possibility of a collision risk decreases, as the target will be more likely to be in a ``non-risky" configuration.
From the table it is also clear that the \gls{KDE} and \gls{DES} based methods yield the same probabilities for each of the simulations.

\subsection{Port Crossing/Head-on scenario}

This scenario contains a \gls{TV} positioned at a relative bearing $\bearing = \ang{354.5}$ and a distance $r = \SI{1000}{\meter}$ from \gls{OS}, such that the initial position is $(N, E) = (995.40, -95.85)$ with course $\course=\ang{174.5}$ and forward velocity $\surgespeed = \SI{10}{\meter\per\second}$.
The scenario is illustrated in Fig.~\ref{fig:sim_port_headon}.
In this scenario the there exists a risk of collision, since $DCPA = \SI{47.98}{\meter} \leq \SI{150}{\meter}$, however the target is placed on the port side of the $\ang{355}$ division head-on/port line by $\ang{0.5}$.
Table \ref{tab:sim_scenario_2} shows the simulation results.
\begin{table*}[tb]
    \centering
    \caption{Scenario 2: Head-on/Port ambiguity}
    \label{tab:sim_scenario_2}
    \begin{tabular}{r|c c c c c c c c c c c c}
        \toprule
         & \multicolumn{2}{c}{$\probability{d_{act}}$} & \multicolumn{2}{c}{$\probability{R_{0}}$} & \multicolumn{2}{c}{$\probability{R_{13}}$} & \multicolumn{2}{c}{$\probability{R_{14}}$} & \multicolumn{2}{c}{$\probability{R_{15}}$} & \multicolumn{2}{c}{$\probability{\text{``give-way"}}$} \\
        \cmidrule(lr){2-3}
        \cmidrule(lr){4-5}
        \cmidrule(lr){6-7}
        \cmidrule(lr){8-9}
        \cmidrule(lr){10-11}
        \cmidrule(lr){12-13}
        $\alpha$ & KDE & DES & KDE & DES & KDE & DES & KDE & DES & KDE & DES & KDE & DES \\
        \midrule
        0.1  &  1.000  &  1.000  &  0.000  &  0.000  &  0.000  &  0.000  &  0.006  &  0.006  &  0.994  &  0.994  &  0.006 &  0.006 \\
        0.5  &  1.000  &  1.000  &  0.000  &  0.000  &  0.000  &  0.000  &  0.337  &  0.336  &  0.663  &  0.664  &  0.337 &  0.336 \\
        1.0  &  0.999  &  1.000  &  0.002  &  0.000  &  0.000  &  0.000  &  0.512  &  0.514  &  0.485  &  0.486  &  0.516 &  0.514 \\
        1.5  &  0.998  &  1.000  &  0.012  &  0.000  &  0.000  &  0.000  &  0.557  &  0.566  &  0.430  &  0.434  &  0.587 &  0.566 \\
        2.0  &  0.993  &  0.994  &  0.023  &  0.003  &  0.000  &  0.000  &  0.551  &  0.569  &  0.425  &  0.428  &  0.610 &  0.570 \\
        5.0  &  0.750  &  0.748  &  0.079  &  0.088  &  0.000  &  0.000  &  0.360  &  0.385  &  0.561  &  0.528  &  0.426 &  0.400 \\
        \bottomrule
    \end{tabular}
\end{table*}
From Table \ref{tab:sim_scenario_2} it is clear that as the uncertainty increases, so does the probability that the situation should be interpreted as a ``head-on", resulting in a higher give-way probability.
It should be noted that this in this scenario higher uncertainty will result in samples that constitutes into $R_0$, or ``no rule".
This is due to the relative bearings getting mapped through \eqref{eq:bearing_colregs_map} to $(PS, PS)$ , which is one of the three cases in Table \ref{tab:colregs_mapping} that cannot be attributed to any of the rules 13-15.
It should also be mentioned that the $\probability{\text{``give-way"}} \approx 0.355 $ for $\alpha = 5$ may at first glance seem high, when compared to the other probabilities, however when the uncertainty is high, some of the samples will be mapped to the $(HO, PS)$, which is a Rule 15 situation with a ``give-way" obligation to ownship, while some other samples gets mapped to $(PS, HO)$, which is a Rule 15 situation with ``stand-on" obligation for ownship.
In other words, the $\probability{R_{15}}$ contains both give-way and stand-on situation in this example.

\subsection{Overtaking/Port Crossing scenario}

This scenario contains a \gls{TV} positioned at a relative bearing $\bearing = \ang{292}$ and a distance $r = \SI{200}{\meter}$ from \gls{OS}, with a course $\course = \ang{0}$ and forward velocity $\surgespeed = \SI{10}{\meter\per\second}$.
In this specific scenario \gls{OS} has a forward velocity of $\surgespeed = \SI{14}{\meter\per\second}$, and a course $\course = \ang{335}$.
This places \gls{OS} $\ang{0.5}$ away from the overtaking/starboard border of the \gls{TV}, in favor of the starboard side (as seen from the \gls{TV}).
The scenario is illustrated in Fig.~\ref{fig:sim_overtaking_starboard}, with the simulation results shown in Table \ref{tab:sim_scenario_3}.
\begin{table*}[tb]
    \centering
    \caption{Scenario 3: Overtaking/Starboard crossing}
    \label{tab:sim_scenario_3}
    \begin{tabular}{r|c c c c c c c c c c c c}
        \toprule
         & \multicolumn{2}{c}{$\probability{d_{act}}$} & \multicolumn{2}{c}{$\probability{R_{0}}$} & \multicolumn{2}{c}{$\probability{R_{13}}$} & \multicolumn{2}{c}{$\probability{R_{14}}$} & \multicolumn{2}{c}{$\probability{R_{15}}$} & \multicolumn{2}{c}{$\probability{\text{``give-way"}}$} \\
        \cmidrule(lr){2-3}
        \cmidrule(lr){4-5}
        \cmidrule(lr){6-7}
        \cmidrule(lr){8-9}
        \cmidrule(lr){10-11}
        \cmidrule(lr){12-13}
        $\alpha$ & KDE & DES & KDE & DES & KDE & DES & KDE & DES & KDE & DES & KDE & DES \\
        \midrule
        0.1 & 0.999 & 1.000 & 0.000 & 0.000 & 0.075 & 0.078 & 0.000 & 0.000 & 0.925 & 0.922 & 0.075 & 0.078 \\
        0.5 & 0.999 & 1.000 & 0.000 & 0.000 & 0.388 & 0.385 & 0.000 & 0.000 & 0.611 & 0.615 & 0.388 & 0.385 \\
        1.0 & 0.996 & 0.997 & 0.000 & 0.000 & 0.460 & 0.444 & 0.000 & 0.000 & 0.558 & 0.556 & 0.440 & 0.442 \\
        1.5 & 0.966 & 0.967 & 0.000 & 0.000 & 0.460 & 0.463 & 0.000 & 0.000 & 0.540 & 0.537 & 0.444 & 0.448 \\
        2.0 & 0.909 & 0.913 & 0.000 & 0.000 & 0.475 & 0.470 & 0.000 & 0.000 & 0.525 & 0.530 & 0.432 & 0.429 \\
        5.0 & 0.625 & 0.624 & 0.001 & 0.000 & 0.489 & 0.488 & 0.000 & 0.000 & 0.510 & 0.512 & 0.306 & 0.304 \\
        \bottomrule
    \end{tabular}
\end{table*}
This scenario illustrates the same increasing and decreasing behaviour in the $\probability{\text{``give-way"}}$ just as in the first scenario, and the explanation is the same: the high uncertainty decreases the chances for an actual collision (as seen in the collision risk $\probability{d_{act}}$).
However, as expected the higher uncertainty leads to more situations evaluated as overtaking, thereby resulting in a higher give-way probability.

\section{Discussion}\label{sec:discussion}

This paper has presented two different methods for implementing and evaluating the uncertainty in \gls{TCPA}, \gls{DCPA} and bearing measurements, and how these uncertainties will affect the evaluation of the different \gls{COLREGs} rules.
As shown in the simulation results, both methods yield comparable results.
However, for both methods the question of how many samples would be sufficient is a valid concern, and will greatly impact the variance of the estimated probabilities, i.e. a higher number of samples results in a lower estimation variance.
The \gls{KDE} based implementation is fast, especially when using the \gls{ISJ} algorithm, and could be even faster if a fixed bandwidth was chosen, however, the bandwidth selection is also the downside of using the \gls{KDE} based method, as this will have a big impact on the accuracy of the density estimate, thus directly affecting the probability estimates.
The bandwidth selection adds another possible inaccuracy to the already existing question of selection of number of samples.

The automata based implementation does require implementing the full understanding automaton with all the logic required, however, the automaton does not require any bandwidth selection, as such it does not suffer from this inaccuracy.
The automata implementation is less sample efficient compared to the \gls{KDE} based method, as all the samples are discarded after being used for selecting the correct transitions in the automaton.
However, as pointed out in Section \ref{sec:method_kde}, the \gls{KDE} memory requirements might not be desirable if implemented in a hardware restricted environment.
The automata implementation can also be parallelized for increased processing speed.

Referring back to the \gls{COLREGs} rules 7 (c) it specifically stated that ``If there is any doubt such risk shall be deemed to exist" \cite{Cockcroft2011}, but how high should the uncertain be before an autonomous system can consider it ``doubt"?
For most statistical tests, a $p$-value limit of $0.05$ is used for a measure of statistical significance, as this covers $2\sigma = 95\%$ of all plausible cases in a normal distribution, could this be compared to  $\probability{\text{``give-way"}} \geq 0.05$ as a limit for ``doubt"?
Looking at the simulation results in Table \ref{tab:sim_scenario_1}, \ref{tab:sim_scenario_2} \& \ref{tab:sim_scenario_3}, selecting $p=0.05$, and using the inequality $\probability{\text{``give-way"}} \geq 0.05$, would result in a ``give-way" action taken by \gls{OS} in all situation except $\alpha = \{0.1, 0.5\}$ in the second scenario (Head-on/port), but is this desirable?
When looking at the configuration of the different scenarios, it is likely that most navigators would err on the safe side and give-way, just to avoid any risks at all.
It can be argued that the limit should be lower than $p=0.25$ such that a scenario where the risk of collision $\probability{d_{act}} = 0.5$, and where two \gls{COLREGs} rules are equally likely with opposing give-way/stand-on obligations, would always lead to a more conservative give-way.
The authors will leave the actual selection of the probability threshold to be implementation specific, however the authors wonder if this threshold is, in some small part, what constitutes ``good seamanship".

\section{Conclusion}\label{sec:conclusion}

This paper presented two probabilistic decision-making methods for evaluating risks of collision in uncertain marine environments, a \gls{KDE} based approach and a \gls{DES} automata based approach.
The methods are based on the standard \gls{TCPA} and \gls{DCPA} calculations, which are the standard method for determining risks of collision in the maritime industry.
The notion of uncertainty is naturally integrated into both methods, enabling a decision-making system to be compliant to \gls{COLREGs} rules 7,13-15 (c), which specifically covers the topic of uncertainty.
Both methods have been evaluated in simulation on three separate edge cases that, in a deterministic setting, might lead to a incorrect decisions, whereas our method is more conservative and decides to give-way in most situations.

Future work will include conditioning on past estimated probabilities, such that a specific situation, e.g. head-on cannot transition to a starboard crossing without subsequent supporting estimates.

\appendix

\subsection{Overview of Relevant COLREGs Rules}\label{app:COLREGS}
This appendix provides a brief list the relevant \gls{COLREGs} rules. See \cite{IMO2021} or \cite{Cockcroft2011} for more details.

\begin{description}
    \item[\emph{Rule 7 - Risk of collision}] \hfill \\ Every vessel shall use all available means appropriate to the prevailing circumstances and conditions to determine if risk of collision exists. If there is any doubt such risk shall be deemed to exist.
    \item[\emph{Rule 13 - Overtaking}] \hfill \\ Any vessel overtaking any other shall keep out of the way of the vessel being overtaken. A vessel shall be deemed to be overtaking when coming up with another vessel from a direction more than 22.5 degrees abaft her beam.
	\item[\emph{Rule 14 - Head-on situation}] \hfill \\ When two power-driven vessels are meeting on nearly reciprocal courses so as to involve risk for collision, then alter course to starboard so that each pass on the port side of each other.
	\item[\emph{Rule 15 - Crossing situation}] \hfill \\ When two power-driven vessels are crossing so as to involve risk of collision, the vessel which has the other on her own starboard side shall keep out of the way and shall, if the circumstances of the case admit, avoid crossing ahead of the other vessel.
	\item[\emph{Rule 16 - Actions by give-way vessel}] \hfill \\ Take early and substantial action to keep well clear.
	\item[\emph{Rule 17 - Actions by stand-on vessel}] \hfill \\ Keep course and speed (be predictable) if possible. If it is necessary to take action, then the ship should try to avoid altering course to port for a vessel on her own port side.
\end{description}

\bibliographystyle{IEEEtran}
\bibliography{bib/references, bib/SLpub}

\vfill

\end{document}